\documentclass[review,hidelinks,onefignum,onetabnum]{siamart251216}

\usepackage{booktabs}
\usepackage{amsmath,amssymb,mathtools,bm}
\newcommand{\norm}[1]{\left\lVert #1\right\rVert}

\usepackage{lipsum}
\usepackage{amsfonts}
\usepackage{graphicx}
\usepackage{epstopdf}
\usepackage{algorithmic}
\ifpdf
  \DeclareGraphicsExtensions{.eps,.pdf,.png,.jpg}
\else
  \DeclareGraphicsExtensions{.eps}
\fi

\newsiamremark{remark}{Remark}
\newsiamremark{hypothesis}{Hypothesis}
\crefname{hypothesis}{Hypothesis}{Hypotheses}
\newsiamthm{claim}{Claim}
\newsiamremark{fact}{Fact}
\crefname{fact}{Fact}{Facts}
\nolinenumbers
\headers{Tracking AI/CI boundary using adjoint equations}{Y. Xiao, H. Li, and Z. Ding}

\title{Tracking the boundary between absolute/convective instability using adjoint equations\thanks{Submitted to the editors DATE.
\funding{Y. Xiao is supported by the National Natural Science Foundation of China
(No.~12402301).
Z. Ding is supported by the National Natural Science Foundation of China (grant nos 12102109, 52176065) and the National Key R\&D programme of China (grant no. 2022YFF0503501)}}}

\author{
Yue Xiao\textsuperscript{1}
\and
Hui Li\textsuperscript{2,3}
\and
Zijing Ding\textsuperscript{2,3,*}
}

\footnotetext[1]{Department of Engineering Mechanics, School of Civil Engineering, Shandong University, Jinan 250061, China}

\footnotetext[2]{School of Civil Engineering, Harbin Institute of Technology, Harbin, Heilongjiang 150001, China}

\footnotetext[3]{International Research Center for Intelligent Fluid Mechanics, Harbin Institute of Technology, Harbin, Heilongjiang 150001, China}

\footnotetext[4]{Corresponding author: \email{z.ding@hit.edu.cn}}

\usepackage{amsopn}

\begin{document}

\maketitle

\begin{abstract}
Determining absolute/convective instability boundaries conventionally requires repeated saddle searches in the complex-wavenumber plane and a subsequent scan of the physical parameter space to locate zero absolute growth. 
Such nested calculations become costly and sensitive to modal branch association for large non-normal eigenvalue problems. 
This work develops a direct continuation method for neutral stationary-saddle boundaries of
frequency-affine generalised eigenvalue problems. 
The zero-group-velocity condition is expressed as an adjoint solvability residual and solved together
with the direct and adjoint eigenproblems, complex gauge constraints and the
neutral-growth condition. 
The resulting one-dimensional solution manifold in the combined parametric space is tracked by scaled pseudo-arclength continuation, allowing parameter folds to be crossed without switching the physical continuation variable.
The formulation recovers the analytical Ginzburg--Landau boundary and, for a Gaussian-wake Orr--Sommerfeld problem, agrees with separately formulated finite-difference saddle corrections to approximately $10^{-8}$ in relative critical Reynolds number. 
Compared with nested complex-wavenumber and parameter-plane saddle scanning, the tested scans require $8.1$--$52.2$ times the wall time of the direct adjoint continuation. 
Extrapolation of the measured cost--accuracy trend to a boundary error of $E_H\sim10^{-6}$ suggests an estimated cost ratio of approximately $1.8\times10^{4}$ in favour of the direct continuation. Application to an Oldroyd--B free-surface film flow reveals genuine folds of the neutral-saddle manifold and a re-entrant CI--AI--CI boundary geometry for the selected saddle family.  
\end{abstract}


\newpage

\section{Introduction}
\label{sec:introduction}

The response of an open flow to a localised disturbance depends not only on
its temporal amplification but also on the rate at which the associated wave
packet is transported away from the source region. 
The distinction between convective and absolute instability is therefore formulated through the
long-time impulse response and the analytic structure of the dispersion
relation in the complex wavenumber--frequency plane
\cite{huerre1985absolute,huerre1990local,chomaz2005global,juniper2014modal}.
In the classical Briggs--Bers framework, an absolute mode is associated with
a branch-point singularity at which spatial branches of opposite spatial
causality pinch, whereas a convectively unstable disturbance is amplified but
swept away from its source region. With the normal-mode convention
\begin{equation}
    \mathbf{q}'(x,y,t)
    =
    \hat{\mathbf{q}}(y)
    \exp\!\left[\mathrm{i}(kx-\omega t)\right],
    \label{eq:normal_mode_convention}
\end{equation}
a neutral absolute/convective instability (AI/CI) boundary is characterised
algebraically by
\begin{equation}
    \frac{\mathrm{d}\omega}{\mathrm{d}k}=0,
    \qquad
    \operatorname{Im}(\omega)=0,
    \label{eq:intro_neutral_saddle}
\end{equation}
provided that the stationary point is the relevant Briggs--Bers pinch
\cite{huerre1985absolute,Schmid2001}. 
Thus, locating an algebraic stationary point and verifying the spatial pinch topology are closely related but logically distinct numerical tasks.

For low-dimensional analytical dispersion relations, the saddle conditions may sometimes be imposed directly. 
In discretised hydrodynamic stability problems, however, the dispersion relation is usually available only
implicitly through a multivalued spectrum. 
A common numerical route is to follow a selected temporal eigenvalue sheet over neighbouring complex
wavenumbers, approximate or interpolate the group velocity, refine a root of 
$\frac{\mathrm{d}\omega}{\mathrm{d}k}=0$, 
and then repeat this saddle calculation while searching the physical parameter space for
$\operatorname{Im}(\omega_s)=0$. 
The numerical difficulties of locating convective/absolute transition points from fully numerical dispersion relations, especially the distinction and continuation of modal branches,
have been analysed in detail by Suslov \cite{suslov2001searching,suslov2006numerical}. 
More recently, Yu and Hunt \cite{yu2023iterative} developed iterative eigenmode tracking to automate
saddle-point calculations for general multivalued dispersion relationships.
Such developments substantially improve the robustness of individual saddle
searches. 
Nevertheless, constructing a complete neutral boundary in a multi-parameter plane still generally requires repeated complex-wavenumber searches or fixed-parameter saddle corrections followed by an outer
zero-growth search or contour reconstruction. 
The cost then multiplies with the number of parameter samples, while inaccurate modal association can
contaminate numerical differentiation when eigenvalue branches approach one another. Alves et al. \cite{alves2019identifying} proposed to use an auxiliary equation derived by differentiating the dispersive relation equation with respect to the wavenumbers to locate the saddle point in the complex wavenumber plane. This method requires no saddle point searching which saves many computational resources. Particularly, the adjoint varibales were used by them for computing the group velocity as well as ensuring solution existence of the auxiliary equation.
Indeed, adjoint equations were used widely for evaluating eigenvalue sensitivity, receptivity, structural sensitivity
\cite{luchini2014adjoint,giannetti2007structural} and
large nonlinear or parameter-dependent eigenproblems \cite{magri2016sensitivity}.  This paper aims to explore its potential in locating the saddle points in the complex wavenumber plane as well as branch tracking that separates the absolute and convective instability regimes, which has been overlooked previously. 


In terms of application and numerical verification,
the proposed formulation is first assessed using two classical problems for which the relevant spatio-temporal stability behaviour is well established.
The linear Ginzburg--Landau equation provides an analytically tractable neutral-saddle boundary and therefore permits a direct verification of the adjoint saddle condition and the pseudo-arclength continuation procedure.
A planar Gaussian wake, a classical model in spatio-temporal wake instability \cite{hultgren1987gaussian}, then provides a non-normal Orr--Sommerfeld eigenvalue problem in which the saddle must be identified on a selected spectral branch. 
And a nested complex-wavenumber/parameter-plane scanning benchmark quantifies the cost of reconstructing the same boundary indirectly. 
These two problems serve as analytical and matrix-level tests, respectively. 
In particular, we consider an Oldroyd--B liquid film flowing underneath an inclined plane. Viscoelasticity is known to have a strong and potentially non-monotonic influence on falling-film stability \cite{shaqfeh1989stability,hu2021stability,jiang2026modelling}, while absolute/convective transitions have been studied for Newtonian inverted films \cite{scheid2016critical} and for other viscoelastic throughflow configurations \cite{hirata2015convective}. 
However, the AI/CI boundary of the present viscoelastic free-surface configuration has not been reported. 
This problem therefore provides a particular application in which the neutral-saddle boundary is not known a priori and may possess a non-trivial dependence on the Reynolds and Deborah numbers.

Overall, the novelty of the present approach lies in formulating the stationary and neutral-growth conditions together with the direct and adjoint eigenproblems as a single augmented algebraic system. 
After one neutral saddle is obtained, the resulting one-dimensional neutral-saddle manifold is tracked
directly by scaled pseudo-arclength continuation \cite{keller1977numerical,mittelmann1986pseudo} 
Consequently, the computation is concentrated on the boundary itself, rather than on sampling a
two-dimensional physical parameter plane and reconstructing the boundary afterwards. 
These features substantially reduce the number of repeated eigenvalue calculations.
The remainder of the paper is organised as follows.
Section~\ref{sec:method} develops the adjoint neutral-saddle formulation,
the gauge-fixed Newton system and the scaled pseudo-arclength continuation
procedure. 
Section~\ref{sec:GL} verifies the formulation using the linear Ginzburg--Landau equation. Section~\ref{sec:wake} considers the planar Gaussian wake, including finite-difference cross-validation, spatial pinch verification and a computational comparison with nested saddle scanning.
Section~\ref{sec:film} applies the method to the Oldroyd--B free-surface film and examines the folded neutral-saddle boundary. 
The principal conclusions and limitations are summarised in Sec.~\ref{sec:conclusions}.

\section{Adjoint formulation and neutral-saddle continuation}
\label{sec:method}

\subsection{Generalised eigenvalue problem and adjoint saddle equation}
\label{subsec:saddle_general}

After spatial discretisation in the transverse direction, the linearised
governing equations and homogeneous boundary conditions are assumed to take
the frequency-affine form
\begin{equation}
    \mathbf{L}(k,\omega;\mathbf{R})\mathbf{q}
    \equiv
    \left[
        \mathbf{A}(k,\mathbf{R})
        -
        \omega\mathbf{B}(k,\mathbf{R})
    \right]\mathbf{q}
    =
    \mathbf{0},
    \label{eq:general_evp}
\end{equation}
where $\mathbf{q}\in\mathbb{C}^{n}$, $k\in\mathbb{C}$,
$\omega\in\mathbb{C}$, and $\mathbf{R}$ is a set of real physical
parameters. A non-trivial solution exists on the dispersion relation
\begin{equation}
    D(k,\omega;\mathbf{R})
    =
    \det\mathbf{L}(k,\omega;\mathbf{R})
    =
    0.
    \label{eq:general_dispersion}
\end{equation}
For a simple finite eigenvalue, Eq.~\eqref{eq:general_dispersion} locally
defines an analytic branch $\omega=\omega(k;\mathbf{R})$.

To express $\frac{\mathrm{d}\omega}{\mathrm{d} k}$ in a form suitable for direct computation we introduce the adjoint problem.
Taking the standard Hermitian inner product $\langle\mathbf{u},\mathbf{v}\rangle=\mathbf{u}^H\mathbf{v}$, where $(\cdot)^H$ denotes the conjugate transpose.
The adjoint eigenvector is defined by
\begin{equation}
    \mathbf{L}^{H}(k,\omega;\mathbf{R})
    \mathbf{q}^{\dagger}
    =
    \mathbf{0}.
    \label{eq:general_adjoint}
\end{equation}

Differentiating Eq.~\eqref{eq:general_evp} with respect to $k$ and taking the inner product with $\mathbf{q}^\dagger$ yields
\begin{equation}\label{eq:deriv}
  \left\langle \mathbf{q}^\dagger, \frac{\partial\mathbf{L}}{\partial k}\mathbf{q} \right\rangle
  +  \frac{\mathrm{d}\omega}{\mathrm{d} k}\left\langle \mathbf{q}^\dagger, \frac{\partial\mathbf{L}}{\partial\omega}\mathbf{q} \right\rangle 
  + \left\langle \mathbf{q}^\dagger, \mathbf{L}\frac{\partial\mathbf{q}}{\partial k} \right\rangle
  = 0.
\end{equation}
The last term vanishes as
$\left\langle \mathbf{q}^\dagger, \mathbf{L}\frac{\partial\mathbf{q}}{\partial k} \right\rangle 
    = \left\langle\frac{\partial\mathbf{q}}{\partial k},\mathbf{L}^H\mathbf{q}^\dagger\right\rangle=0$. 
Since $\frac{\partial\mathbf{L}}{\partial\omega} = -\mathbf{B}$, Eq. (\ref{eq:deriv}) is restated as
\begin{equation}
    \frac{\mathrm{d}\omega}{\mathrm{d} k}
    =
    \frac{
    (\mathbf{q}^{\dagger})^H
    \left(
        \frac{\partial\mathbf{A}}{\partial k}
        -
        \omega\frac{\partial\mathbf{B}}{\partial k}
    \right)
    \mathbf{q}
    }{
    (\mathbf{q}^{\dagger})^H
    \mathbf{B}\mathbf{q}
    }.
    \label{eq:adjoint_sensitivity}
\end{equation}
For a simple finite eigenvalue,
$(\mathbf{q}^{\dagger})^H\mathbf{B}\mathbf{q}\neq0$. 
Defining $    \mathbf{C}(k,\omega;\mathbf{R})
    =
    \frac{\partial\mathbf{A}}{\partial k}
    -
    \omega\frac{\partial\mathbf{B}}{\partial k},$
the stationary-point condition is equivalent to the complex algebraic
equation
\begin{equation}
    \mathcal{S}
    \equiv
    (\mathbf{q}^{\dagger})^H
    \mathbf{C}\mathbf{q}
    =
    0.
    \label{eq:adjoint_saddle_condition}
\end{equation}
The sign in Eq.~\eqref{eq:adjoint_sensitivity} follows directly from the
definition $\mathbf{L}=\mathbf{A}-\omega\mathbf{B}$.

\subsection{Gauge-fixed Newton system for a saddle at fixed parameters}
\label{subsec:saddle_newton}

At fixed $\mathbf{R}$, the unknown saddle state is
$(\mathbf{q},\mathbf{q}^{\dagger},\omega,k)$. The direct and adjoint
eigenvectors possess independent complex scale and phase freedoms. Merely
adding the biorthogonality condition
$(\mathbf{q}^{\dagger})^H\mathbf{B}\mathbf{q}=1$ does not remove the
combined transformation
\begin{equation}
    \mathbf{q}\mapsto c\mathbf{q},
    \qquad
    \mathbf{q}^{\dagger}\mapsto
    \frac{\mathbf{q}^{\dagger}}{\overline{c}},
    \qquad c\in\mathbb{C}\setminus\{0\}.
    \label{eq:combined_gauge}
\end{equation}
A regular Newton system therefore requires two independent complex gauge
conditions.

Let $\mathbf{q}_{\mathrm{ref}}$ be a fixed reference vector within one
Newton corrector and impose $\mathbf{q}_{\mathrm{ref}}^H\mathbf{q}=1.$
The adjoint scale is fixed by
\begin{equation}
    (\mathbf{q}^{\dagger})^H\mathbf{B}\mathbf{q}=1.
    \label{eq:adjoint_gauge}
\end{equation}
Because $\mathbf{L}\mathbf{q}=\mathbf{0}$ implies that the rows of
$\mathbf{L}^H$ are linearly dependent at a converged eigenpair, one complex
adjoint equation is redundant. Let
$\mathbf{P}_j\in\mathbb{R}^{(n-1)\times n}$ delete the $j$th adjoint
residual component. 
In the implementation, $j=\arg\max_{\ell}|q_{\ell}|$
is used to avoid deleting a row associated with a negligibly small
coefficient in the row dependence.

The fixed-parameter saddle is obtained from
\begin{equation}
    \mathcal{F}_{s}
    =
    \begin{pmatrix}
        \mathbf{L}\mathbf{q}\\[2pt]
        \mathbf{P}_{j}\mathbf{L}^{H}\mathbf{q}^{\dagger}\\[2pt]
        (\mathbf{q}^{\dagger})^H\mathbf{C}\mathbf{q}\\[2pt]
        \mathbf{q}_{\mathrm{ref}}^H\mathbf{q}-1\\[2pt]
        (\mathbf{q}^{\dagger})^H\mathbf{B}\mathbf{q}-1
    \end{pmatrix}
    =
    \mathbf{0}.
    \label{eq:fixed_saddle_system}
\end{equation}
Equation~\eqref{eq:fixed_saddle_system} contains
$2n+2$ complex equations for $2n+2$ complex unknowns.
The corresponding real state is
\begin{equation}
    \mathbf{U}
    =
    \left(
    \mathbf{q}_{r}^{T},
    \mathbf{q}_{i}^{T},
    (\mathbf{q}^{\dagger}_{r})^{T},
    (\mathbf{q}^{\dagger}_{i})^{T},
    \omega_r,\omega_i,k_r,k_i
    \right)^T
    \in\mathbb{R}^{4n+4}.
    \label{eq:fixed_saddle_real_state}
\end{equation}
The real and imaginary parts of Eq.~\eqref{eq:fixed_saddle_system} form a
square real system $\mathbf{F}_{s}(\mathbf{U})=\mathbf{0}$, solved by
damped Newton iteration,
\begin{equation}
    \mathbf{J}_{s}(\mathbf{U}^{(r)})
    \Delta\mathbf{U}^{(r)}
    =
    -\mathbf{F}_{s}(\mathbf{U}^{(r)}),
    \qquad
    \mathbf{U}^{(r+1)}
    =
    \mathbf{U}^{(r)}
    +
    \gamma_r\Delta\mathbf{U}^{(r)},
    \label{eq:fixed_saddle_newton}
\end{equation}
where $0<\gamma_r\leq1$ is selected by a residual-decrease line search.

The wavenumber derivatives of the operators $\mathbf{A}$ and $\mathbf{B}$ can be evaluated either
analytically, by direct differentiation of their closed‑form expressions, or numerically by centred
finite differences. Each approach has its own merits.
An analytical derivative, e.g.\ $\frac{\partial\mathbf{A}}{\partial k}$ obtained by differentiating the operator
expression with respect to $k$, is exact (up to working precision) and free of discretisation error.
However, it requires the operators to be available in a differentiable form, and the algebra may
become cumbersome for complicated operators.  
The numerical alternative is simple to implement and only needs the operator evaluations
\begin{align}
    \frac{\partial\mathbf{A}}{\partial k}
    &\simeq
    \frac{
        \mathbf{A}(k+h_k)-\mathbf{A}(k-h_k)
    }{2h_k},
    \label{eq:dA_dk_fd}
    \\
    \frac{\partial\mathbf{B}}{\partial k}
    &\simeq
    \frac{
        \mathbf{B}(k+h_k)-\mathbf{B}(k-h_k)
    }{2h_k},
    \label{eq:dB_dk_fd}
\end{align}
with a real increment $h_k$. This centred‑difference formula yields an $\mathcal{O}(h_k^2)$
approximation to the complex derivative with respect to $k$ when the discrete operators are
analytic in a neighbourhood of the saddle. The step size $h_k$ must be chosen to balance
truncation error and round‑off error, whereas the analytical route avoids this trade‑off entirely.

An initial state is constructed from an approximate pair
$(k_{\mathrm g},\omega_{\mathrm g})$. At $k=k_{\mathrm g}$, the generalised
temporal eigenproblem is solved and the finite eigenvalue closest to
$\omega_{\mathrm g}$ is selected. Its right eigenvector defines
$\mathbf{q}$ and the local reference gauge. The initial adjoint vector is
the right singular vector corresponding to the smallest singular value of
$\mathbf{L}^H$ and is rescaled to satisfy Eq.~\eqref{eq:adjoint_gauge}.

\subsection{Neutral-saddle manifold in a two-parameter plane}
\label{subsec:aici_extended}

We now choose two real control parameters
$\mathbf{R}=(R_1,R_2)$ and hold all other parameters fixed. The AI/CI
boundary is represented algebraically by the neutral-saddle equations
\begin{equation}
    \frac{\mathrm{d}\omega}{\mathrm{d} k}=0,
    \qquad
    \omega_i\equiv\operatorname{Im}(\omega)=0.
    \label{eq:neutral_saddle_conditions}
\end{equation}
Using Eq.~\eqref{eq:adjoint_saddle_condition}, the gauge-fixed boundary
residual is
\begin{equation}
    \mathcal{F}_{b}
    (\mathbf{q},\mathbf{q}^{\dagger},\omega,k,R_1,R_2)
    =
    \begin{pmatrix}
        \mathbf{L}\mathbf{q}\\[2pt]
        \mathbf{P}_{j}\mathbf{L}^{H}\mathbf{q}^{\dagger}\\[2pt]
        (\mathbf{q}^{\dagger})^H\mathbf{C}\mathbf{q}\\[2pt]
        \mathbf{q}_{\mathrm{ref}}^H\mathbf{q}-1\\[2pt]
        (\mathbf{q}^{\dagger})^H\mathbf{B}\mathbf{q}-1\\[2pt]
        \omega_i
    \end{pmatrix}
    =
    \mathbf{0}.
    \label{eq:boundary_system_complex}
\end{equation}
In fact, by solving Eq.~(\ref{eq:boundary_system_complex}) we can obtain contours along which $\omega_i$ takes any prescribed value.
Separating complex equations into real and imaginary parts gives
$4n+5$ real equations. The continuation state
\begin{equation}
    \mathbf{X}
    =
    \left(
    \mathbf{q}_{r}^{T},
    \mathbf{q}_{i}^{T},
    (\mathbf{q}^{\dagger}_{r})^{T},
    (\mathbf{q}^{\dagger}_{i})^{T},
    \omega_r,\omega_i,k_r,k_i,R_1,R_2
    \right)^T
    \in\mathbb{R}^{4n+6}
    \label{eq:continuation_state}
\end{equation}
therefore satisfies an underdetermined real system
\begin{equation}
    \mathbf{F}_{b}(\mathbf{X})=\mathbf{0},
    \qquad
    \mathbf{F}_{b}:
    \mathbb{R}^{4n+6}
    \rightarrow
    \mathbb{R}^{4n+5}.
    \label{eq:boundary_real_system}
\end{equation}
At a regular solution, Eq.~\eqref{eq:boundary_real_system} defines a
one-dimensional manifold in state--parameter  space. Its projection onto the
$(R_1,R_2)$ plane is the algebraic neutral-saddle boundary.

\subsection{Scaled pseudo-arclength predictor--corrector}
\label{subsec:aici_palc}

Natural continuation in $R_1$ or $R_2$ fails when the selected parameter
ceases to be a valid local coordinate, for example when
$\mathrm{d} R_2/\mathrm{d} s=0$ on the solution curve. 
To avoid manual parameter switching, continuation is performed in scaled coordinates
\begin{equation}
    \mathbf{X}=\mathbf{S}_x\mathbf{Z},
    \label{eq:state_scaling}
\end{equation}
where $\mathbf{S}_x$ is a fixed positive diagonal scaling matrix constructed
once from the corrected initial state $\mathbf{X}_0$. The scaling is assigned
blockwise,
\begin{equation}
    \mathbf{S}_x
    =
    \operatorname{diag}
    \left(
        d_q\mathbf{I}_{2n},
        d_{q^\dagger}\mathbf{I}_{2n},
        d_\omega\mathbf{I}_{2},
        d_k\mathbf{I}_{2},
        d_{R_1},
        d_{R_2}
    \right),
    \label{eq:state_scaling_blocks}
\end{equation}
so that the real and imaginary parts of each complex quantity share the same
scale. The scalar scales are formed from the initial corrected state according
to
\begin{equation}
    d_\omega=\max(1,|\omega_0|),
    \qquad
    d_k=\max(1,|k_0|),
    \qquad
    d_{R_j}=\max(1,|R_{j,0}|),
    \quad j=1,2,
    \label{eq:state_scaling_scalar_rule}
\end{equation}
while $d_q$ and $d_{q^\dagger}$ are fixed block scales determined from the
magnitudes of the corresponding initial direct and adjoint modes, with a
positive lower bound used to avoid a vanishing scale. Once constructed,
$\mathbf{S}_x$ is held fixed throughout both directional continuation traces.
The scaled residual and Jacobian are
\begin{equation}
    \widehat{\mathbf{F}}_{b}(\mathbf{Z})
    =
    \mathbf{F}_{b}(\mathbf{S}_x\mathbf{Z}),
    \qquad
    \mathbf{J}_{Z}
    =
    \mathbf{J}_{X}\mathbf{S}_{x}.
    \label{eq:scaled_jacobian}
\end{equation}

At a regular point $\mathbf{Z}_m$,
$\mathbf{J}_{Z}\in\mathbb{R}^{(4n+5)\times(4n+6)}$ has a one-dimensional
right nullspace. The unit tangent in scaled state space is denoted by
$\mathbf{t}_m^{(z)}$ and satisfies
\begin{equation}
    \mathbf{J}_{Z}(\mathbf{Z}_m)\mathbf{t}_m^{(z)}
    =
    \mathbf{0},
    \qquad
    \norm{\mathbf{t}_m^{(z)}}_2=1.
    \label{eq:tangent_equation}
\end{equation}
The corresponding tangent in the original state coordinates is
\begin{equation}
    \mathbf{t}_m^{(x)}
    =
    \mathbf{S}_x\mathbf{t}_m^{(z)}.
    \label{eq:physical_state_tangent}
\end{equation}
For geometric diagnostics in the physical $(R_1,R_2)$ parameter plane, we
further define the normalised parameter tangent
\begin{equation}
    \mathbf{t}_m^{(p)}
    =
    \frac{
        \begin{pmatrix}
            (\mathbf{t}_m^{(x)})_{R_1}\\
            (\mathbf{t}_m^{(x)})_{R_2}
        \end{pmatrix}
    }{
        \left[
            ((\mathbf{t}_m^{(x)})_{R_1})^2
            +
            ((\mathbf{t}_m^{(x)})_{R_2})^2
        \right]^{1/2}
    }
    =
    \begin{pmatrix}
        t_{R_1}^{(p)}\\
        t_{R_2}^{(p)}
    \end{pmatrix}.
    \label{eq:parameter_plane_tangent}
\end{equation}
Thus, $\mathbf{t}^{(z)}$ is used in the scaled pseudo-arclength
predictor--corrector, whereas $\mathbf{t}^{(p)}$ is used to identify turning
points and describe the geometry of the projected boundary in physical
parameter space.

The implementation computes the full right-singular-vector matrix of
$\mathbf{J}_{Z}$ and selects the final right singular vector. An
economy-size singular value decomposition must not be used for this wide
$m\times(m+1)$ matrix because it omits the additional right-null vector.
For $m>0$, the orientation is chosen such that
\begin{equation}
    (\mathbf{t}_{m}^{(z)})^{T}\mathbf{t}_{m-1}^{(z)}>0.
\end{equation}

An Euler predictor is
\begin{equation}
    \mathbf{Z}_{m+1}^{p}
    =
    \mathbf{Z}_{m}
    +
    \Delta s\,\mathbf{t}_{m}^{(z)}.
\end{equation}
The corrector solves
\begin{equation}
    \mathbf{G}(\mathbf{Z})
    =
    \begin{pmatrix}
        \widehat{\mathbf{F}}_{b}(\mathbf{Z})\\[2pt]
        (\mathbf{t}_{m}^{(z)})^{T}
        \left(
        \mathbf{Z}-\mathbf{Z}_{m+1}^{p}
        \right)
    \end{pmatrix}
    =
    \mathbf{0}.
    \label{eq:palc_corrector}
\end{equation}
At Newton iteration $r$,
\begin{equation}
    \begin{bmatrix}
        \mathbf{J}_{Z}(\mathbf{Z}^{(r)})\\
        (\mathbf{t}_{m}^{(z)})^{T}
    \end{bmatrix}
    \Delta\mathbf{Z}^{(r)}
    =
    -
    \begin{pmatrix}
        \widehat{\mathbf{F}}_{b}(\mathbf{Z}^{(r)})\\
        (\mathbf{t}_{m}^{(z)})^{T}
        \left(
        \mathbf{Z}^{(r)}-\mathbf{Z}_{m+1}^{p}
        \right)
    \end{pmatrix}.
    \label{eq:palc_newton}
\end{equation}

One physical parameter is fixed only during initial neutral-saddle
correction. For example,
\begin{equation}
    \begin{pmatrix}
        \mathbf{F}_{b}(\mathbf{X})\\
        R_2-R_{2,\mathrm g}
    \end{pmatrix}
    =
    \mathbf{0}
    \label{eq:initial_fixed_parameter}
\end{equation}
is a square system that corrects a nearby saddle to the AI/CI boundary.
After this first point converges, the fixed-parameter equation is removed,
the scaled nullspace tangent is calculated, and the two orientations
$\pm\mathbf{t}_0^{(z)}$ are traced separately. The arclength step is increased
when the corrector converges rapidly and reduced when Newton convergence is
slow or a trial step is rejected.

\begin{algorithm}
\caption{Adjoint pseudo-arclength continuation of a neutral-saddle branch}
\label{alg:adjoint_palc}
\begin{algorithmic}

\STATE{Given a nearby saddle estimate, fixed non-continuation parameters,
and an initial arclength step $\Delta s$}

\STATE{Select the temporal eigenvalue closest to the supplied frequency
estimate and construct the corresponding direct eigenvector $\mathbf{q}$}

\STATE{Construct the adjoint eigenvector
$\mathbf{q}^{\dagger}$ from the smallest singular vector of
$\mathbf{L}^{H}$ and impose the biorthogonal gauge}

\STATE{Fix one physical parameter and solve the augmented neutral-saddle
system
$\mathbf{F}_{b}(\mathbf{X})=\mathbf{0}$
to obtain the initial boundary point
$\mathbf{X}_{0}$}

\STATE{Construct the fixed scaling matrix
$\mathbf{S}_{x}$ and set
$\mathbf{Z}_{0}=\mathbf{S}_{x}^{-1}\mathbf{X}_{0}$}

\STATE{Compute the scaled tangent
$\mathbf{t}_{0}^{(z)}$
from the null space of
$\mathbf{J}_{Z}(\mathbf{Z}_{0})$ and normalise it}

\FOR{each initial orientation
$\mathbf{t}_{0}^{(z)}$ and $-\mathbf{t}_{0}^{(z)}$}

    \WHILE{the prescribed parameter window has not been exhausted}

        \STATE{Predict the next point:
        $\mathbf{Z}_{m+1}^{p}
        =
        \mathbf{Z}_{m}
        +
        \Delta s\,\mathbf{t}_{m}^{(z)}$}

        \STATE{Correct the predictor by solving the pseudo-arclength
        system
        $\mathbf{G}(\mathbf{Z})=\mathbf{0}$
        with damped Newton iteration}

        \IF{the Newton corrector fails}

            \STATE{Reduce $\Delta s$ and repeat the predictor--corrector step}

        \ENDIF

        \STATE{Update the local direct-mode gauge and deleted adjoint
        residual component}

        \STATE{Compute and orient the new scaled tangent
        $\mathbf{t}_{m+1}^{(z)}$}

        \STATE{Evaluate continuation diagnostics and store the accepted
        boundary point}

        \STATE{Adapt $\Delta s$ according to the Newton convergence
        behaviour}

    \ENDWHILE

\ENDFOR

\STATE{Merge the two directional traces into a single connected
neutral-saddle branch}

\RETURN{Neutral-saddle boundary branch}

\end{algorithmic}
\end{algorithm}

\subsection{Semi-analytical Jacobian and numerical diagnostics}
\label{subsec:semianalytic_jacobian}

A centred-difference approximation of every Jacobian column is inefficient
because $4n$ real unknowns are direct- and adjoint-mode components. At fixed
$(\omega,k,R_1,R_2)$, the eigenvector blocks can be assembled exactly from
\begin{align}
    \delta(\mathbf{L}\mathbf{q})
    &=
    \mathbf{L}\,\delta\mathbf{q},
    \label{eq:jac_direct_variation}\\
    \delta(\mathbf{P}_j\mathbf{L}^{H}\mathbf{q}^{\dagger})
    &=
    \mathbf{P}_j\mathbf{L}^{H}\delta\mathbf{q}^{\dagger},
    \label{eq:jac_adjoint_variation}\\
    \delta\mathcal{S}
    &=
    (\mathbf{q}^{\dagger})^H\mathbf{C}\,\delta\mathbf{q}
    +
    (\delta\mathbf{q}^{\dagger})^H\mathbf{C}\mathbf{q},
    \label{eq:jac_saddle_variation}\\
    \delta(\mathbf{q}_{\mathrm{ref}}^H\mathbf{q})
    &=
    \mathbf{q}_{\mathrm{ref}}^H\delta\mathbf{q},
    \label{eq:jac_direct_gauge_variation}\\
    \delta\!\left[
    (\mathbf{q}^{\dagger})^H\mathbf{B}\mathbf{q}
    \right]
    &=
    (\mathbf{q}^{\dagger})^H\mathbf{B}\,\delta\mathbf{q}
    +
    (\delta\mathbf{q}^{\dagger})^H\mathbf{B}\mathbf{q}.
    \label{eq:jac_adjoint_gauge_variation}
\end{align}
After conversion to real blocks, Eqs.~\eqref{eq:jac_direct_variation}--
\eqref{eq:jac_adjoint_gauge_variation} fill the $4n$ eigenvector columns
analytically. Only the six scalar columns
$(\omega_r,\omega_i,k_r,k_i,R_1,R_2)$ are approximated by centred
differences. The same Jacobian assembly is used for the tangent calculation
and the pseudo-arclength corrector.

At each accepted point, the implementation monitors
\begin{equation}
\begin{split}
    r_q &= \norm{\mathbf{L}\mathbf{q}}_2,\\
    r_{q^\dagger} &= \norm{\mathbf{L}^{H}\mathbf{q}^{\dagger}}_2,\\
    r_s &= |(\mathbf{q}^{\dagger})^H\mathbf{C}\mathbf{q}|,\\
    r_{\omega_k} &= \left|\frac{\mathrm{d}\omega}{\mathrm{d} k}\right|,\\
    r_n &= |\omega_i|,\\
    r_b &= |(\mathbf{q}^{\dagger})^H\mathbf{B}\mathbf{q}-1|,\\
    r_t &= \frac{\norm{\mathbf{J}_{Z}\mathbf{t}^{(z)}}_2}
    {\max(1,\norm{\mathbf{J}_{Z}}_{\mathrm F})}.
\end{split}
\label{eq:continuation_diagnostics}
\end{equation}
The singular values of $\mathbf{J}_{Z}$ are additionally recorded as a rank
diagnostic near folds and possible mode interactions.

Equations~\eqref{eq:adjoint_saddle_condition} and
\eqref{eq:neutral_saddle_conditions} identify algebraic stationary
eigenpairs. They do not by themselves prove that a stationary point is a
physical pinch. The initial seed must therefore be checked by spatial branch
tracking according to the Briggs--Bers criterion\cite{Schmid2001}. Representative points are
rechecked near parameter folds, rapid changes in the continued eigenmode, or
a deterioration of the Jacobian rank indicator. This separation between
algebraic continuation and physical pinch verification is maintained
throughout the paper.

\section{Application to the Ginzburg--Landau equation}
\label{sec:GL}

The linear Ginzburg--Landau equation provides a scalar problem for which the
neutral-saddle boundary, the eigenvalue derivative, and the continued saddle
coordinates are available analytically. It therefore tests whether the general
direct--adjoint formulation of Sec.~\ref{sec:method} reduces correctly in the
one-dimensional limit, before the method is applied to matrix eigenvalue
problems.

Consider the linear advection--diffusion equation
\begin{equation}
    \frac{\partial q}{\partial t}
    +
    U\frac{\partial q}{\partial x}
    =
    \mu q
    +
    D\frac{\partial^2 q}{\partial x^2},
    \qquad D>0,
    \label{eq:GL_equation}
\end{equation}
where $U$ is the advection velocity, $\mu$ is the temporal amplification
parameter, and $D$ is the diffusion coefficient. With the normal-mode
convention in Eq.~\eqref{eq:normal_mode_convention},
$q=\hat q\exp[\mathrm{i}(kx-\omega t)]$, Eq.~\eqref{eq:GL_equation} gives
\begin{equation}
    \omega(k)
    =
    Uk+\mathrm{i}(\mu-Dk^2).
    \label{eq:GL_dispersion}
\end{equation}

\subsection{Scalar direct--adjoint formulation}
\label{subsec:GL_adjoint}

To apply exactly the same notation as in Sec.~\ref{subsec:saddle_general},
Eq.~\eqref{eq:GL_dispersion} is written as the scalar frequency-affine
eigenproblem
\begin{equation}
    L_{\mathrm{GL}}q
    \equiv
    \left[
        A_{\mathrm{GL}}(k,U,\mu)
        -
        \omega B_{\mathrm{GL}}
    \right]q
    =0,
    \label{eq:GL_scalar_evp}
\end{equation}
with $A_{\mathrm{GL}} = Uk+\mathrm{i}(\mu-Dk^2)$ and $B_{\mathrm{GL}}=1.$
The scalar adjoint problem is
\begin{equation}
    L_{\mathrm{GL}}^{H}q^{\dagger}=0.
    \label{eq:GL_adjoint}
\end{equation}
The wavenumber derivative entering the adjoint saddle equation is
\begin{equation}
    C_{\mathrm{GL}}
    \equiv
    \frac{\partial A_{\mathrm{GL}}}{\partial k}
    -
    \omega
    \frac{\partial B_{\mathrm{GL}}}{\partial k}
    =
    U-2\mathrm{i} Dk.
    \label{eq:GL_C}
\end{equation}
Hence the adjoint sensitivity identity~\eqref{eq:adjoint_sensitivity}
becomes
\begin{equation}
    \frac{\mathrm{d}\omega}{\mathrm{d} k}
    =
    \frac{
        (q^{\dagger})^{H}
        C_{\mathrm{GL}}q
    }{
        (q^{\dagger})^{H}
        B_{\mathrm{GL}}q
    }
    =
    U-2\mathrm{i} Dk,
    \label{eq:GL_adjoint_derivative}
\end{equation}
which is identical to the analytical derivative of
Eq.~\eqref{eq:GL_dispersion}.

For $n=1$, deleting one redundant adjoint equation leaves an empty reduced
adjoint block, because
$\mathbf{P}_{j}\in\mathbb{R}^{0\times1}$. The adjoint variable is nevertheless
retained in the saddle residual and in the biorthogonal gauge. With
$q_{\mathrm{ref}}=1$, the scalar neutral-saddle residual corresponding to
Eq.~\eqref{eq:boundary_system_complex} is
\begin{equation}
    \mathcal{F}_{\mathrm{GL}}
    =
    \begin{pmatrix}
        L_{\mathrm{GL}}q\\[2pt]
        (q^{\dagger})^H C_{\mathrm{GL}}q\\[2pt]
        q-1\\[2pt]
        (q^{\dagger})^Hq-1\\[2pt]
        \omega_i
    \end{pmatrix}
    =
    \mathbf{0}.
    \label{eq:GL_extended_system}
\end{equation}
The first four blocks in Eq.~\eqref{eq:GL_extended_system} are complex and the
last equation is real. Thus the residual supplies nine real equations for the
ten-dimensional real continuation state
\begin{equation}
    \mathbf{X}_{\mathrm{GL}}
    =
    \left(
        q_r,q_i,
        q_r^{\dagger},q_i^{\dagger},
        \omega_r,\omega_i,
        k_r,k_i,
        U,\mu
    \right)^T
    \in\mathbb{R}^{10}.
    \label{eq:GL_real_state}
\end{equation}
The two complex gauge equations in Eq.~\eqref{eq:GL_extended_system} give
$q=q^{\dagger}=1$. Consequently, the adjoint saddle equation reduces to
$C_{\mathrm{GL}} = U-2\mathrm{i} Dk =0,$
and therefore
$k_s=-\frac{\mathrm{i} U}{2D}.$
Substitution into Eq.~\eqref{eq:GL_dispersion} yields
$\omega_s =\mathrm{i}\left(\mu-\frac{U^2}{4D}\right).$
The neutral-saddle condition $\omega_i=0$ then gives the analytical boundary
\begin{equation}
    \mu_c=\frac{U^2}{4D}.
    \label{eq:GL_boundary}
\end{equation}
Thus, in the scalar limit, the general adjoint residual does not merely recover
the same saddle coordinates as direct differentiation; it reduces identically
to the exact group-velocity condition.

For completeness, the exact first variations used to assemble the real Newton
Jacobian in this verification problem are
\begin{equation}
    \delta L_{\mathrm{GL}}
    =
    k\,\delta U
    +
    C_{\mathrm{GL}}\,\delta k
    +
    \mathrm{i}\,\delta\mu
    -
    \delta\omega,
    \qquad
    \delta C_{\mathrm{GL}}
    =
    \delta U-2\mathrm{i} D\,\delta k.
    \label{eq:GL_operator_variations}
\end{equation}
These expressions are inserted into the variation formulas
\eqref{eq:jac_direct_variation}--\eqref{eq:jac_adjoint_gauge_variation}, so
that an exact real Jacobian is used rather than a finite-difference
Jacobian.

\begin{figure}[htbp]
    \centering
    \includegraphics[width=0.6\textwidth]{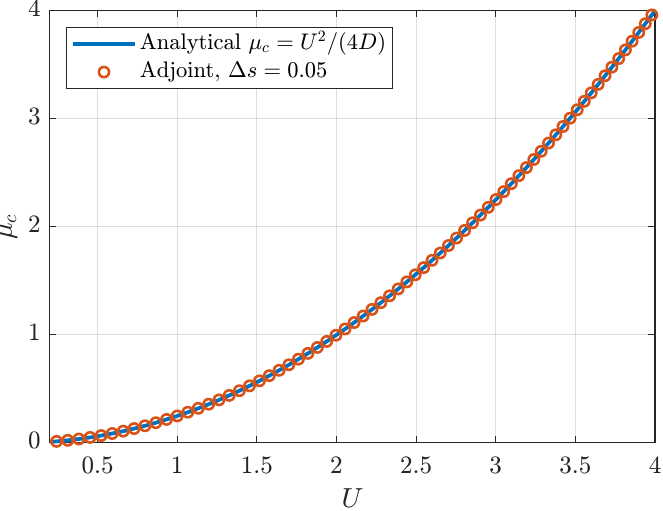}
    \caption{Analytical neutral-saddle boundary
    $\mu_c=U^2/(4D)$ and the boundary obtained from the gauge-fixed
    direct--adjoint pseudo-arclength formulation for the linear
    Ginzburg--Landau equation.}
    \label{fig:GL_boundary}
\end{figure}

\subsection{Numerical continuation and error measures}
\label{subsec:GL_results}

The numerical test fixes $D=1$ and treats $(U,\mu)$ as the two continuation
parameters. A deliberately perturbed initial estimate at $U=1$ is first
corrected to a neutral saddle by solving
Eq.~\eqref{eq:GL_extended_system} together with the fixed-parameter condition
$U=1$. The fixed-$U$ equation is then removed and the resulting $9\times10$
real Jacobian is used to compute the one-dimensional right nullspace tangent.
The two tangent orientations are followed with the scaled pseudo-arclength
predictor--corrector algorithm of Sec.~\ref{subsec:aici_palc} over
$0.2\leq U\leq4$. Three fixed arclength steps,
$\Delta s=0.20$, $0.10$, and $0.05$, are considered.

Two independent checks are performed. First, the adjoint derivative
Eq.~\eqref{eq:GL_adjoint_derivative} is evaluated at six non-stationary complex
wavenumbers and compared with the exact derivative $U-2\mathrm{i} Dk$. Evaluating the
identity away from the saddle avoids the trivial comparison of two quantities
that both vanish. Second, the continued neutral-saddle branch is compared
pointwise with Eq.~\eqref{eq:GL_boundary}. The boundary errors are defined as
\begin{equation}
    E_{\infty}
    =
    \max_m
    \left|
        \mu_m-\frac{U_m^2}{4D}
    \right|,
    \qquad
    E_{\mathrm{RMS}}
    =
    \left[
        \frac{1}{N_p}
        \sum_{m=1}^{N_p}
        \left(
            \mu_m-\frac{U_m^2}{4D}
        \right)^2
    \right]^{1/2}.
    \label{eq:GL_error_measures}
\end{equation}

For the six off-saddle derivative checks, the maximum absolute discrepancy
between the adjoint and analytical derivatives is
$0$, with a maximum scaled relative discrepancy of
$0$. For the finest continuation step
$\Delta s=0.05$, 69 boundary points are accepted. The
maximum and RMS errors relative to the analytical boundary are
$8.882\times10^{-16}$ and $2.610\times10^{-16}$, respectively. The maximum saddle,
neutral-growth, and tangent-nullspace residuals are
$0$, $1.913\times10^{-24}$, and $1.307\times10^{-16}$. 
The three-step comparison in Table~\ref{tab:GL_validation}  confirms that the nonlinear corrector returns predicted states to the analytical neutral-saddle manifold over a range of arclength steps.

\begin{table}[htbp]
    \centering
    \caption{Pseudo-arclength verification for the linear Ginzburg--Landau equation.}
    \label{tab:GL_validation}
    \begin{tabular}{cccccc}
        \toprule
        $\Delta s$ & $N_p$ & $E_{\infty}$ & $E_{\mathrm{RMS}}$ & $\max r_s$ & $\max r_n$ \\
        \midrule
        0.2 & 17 & 5.502e-13 & 3.089e-13 & 0.000e+00 & 4.946e-22 \\
        0.1 & 34 & 2.442e-15 & 1.323e-15 & 0.000e+00 & 3.684e-23 \\
        0.05 & 69 & 8.882e-16 & 2.610e-16 & 0.000e+00 & 1.913e-24 \\
        \bottomrule
    \end{tabular}
\end{table}

Figure~\ref{fig:GL_boundary} provides a direct scalar verification of the full
algorithmic sequence used later for the matrix problems: neutral-saddle
correction, nullspace-tangent evaluation, two-sided pseudo-arclength
continuation, and residual-based acceptance. Because the analytical branch is
known exactly, any discrepancy in this test originates from the nonlinear
continuation implementation rather than from spatial discretisation or modal
branch association.

\section{Application to a planar Gaussian wake}
\label{sec:wake}

The linear Ginzburg--Landau problem verifies the scalar limit of the
direct--adjoint formulation. We next consider a matrix-valued, non-normal
eigenvalue problem for which the saddle condition must be evaluated on a
selected eigenvalue branch. A planar Gaussian wake is used for this purpose.
The same family of velocity profiles has been widely employed in
spatio-temporal wake analysis~\cite{hultgren1987gaussian}. In the present
study, the wake problem is used to assess the mapped spectral discretisation,
the pseudo-arclength boundary continuation, an independently formulated
finite difference neutral-saddle cross validation, and the identification of
the corresponding spatial pinch topology.

\subsection{Orr--Sommerfeld formulation and mapped collocation}
\label{subsec:wake_formulation}

The parallel base flow is
\begin{equation}
    U(y)
    =
    1-\lambda\exp[-(\ln2)y^2],
    \label{eq:wake_profile}
\end{equation}
where $\lambda$ denotes the centreline velocity-deficit parameter and the
wake half-width is used as the reference length. For a two-dimensional
disturbance
\begin{equation}
    \psi(x,y,t)
    =
    \phi(y)\exp[\mathrm{i}(\alpha x-\omega t)],
\end{equation}
the linearised equations reduce to the Orr--Sommerfeld equation
\begin{equation}
    (U-c)(D^2-\alpha^2)\phi
    -
    U''\phi
    =
    -\frac{\mathrm{i}}{\alpha Re}
    (D^2-\alpha^2)^2\phi,
    \qquad
    c=\frac{\omega}{\alpha},
    \label{eq:wake_OS}
\end{equation}
with far-field conditions
\begin{equation}
    \phi(\pm\infty)
    =
    D\phi(\pm\infty)
    =
    0.
    \label{eq:wake_farfield}
\end{equation}

A direct linear map from the Chebyshev interval to a large truncated
physical domain is found to resolve the narrow wake core inefficiently.
The production calculations therefore use the sinh map
\begin{equation}
    y(x)
    =
    y_{\max}
    \frac{\sinh(\beta x)}{\sinh\beta},
    \qquad
    -1\le x\le1,
    \label{eq:wake_sinh_map}
\end{equation}
where $\beta\ge0$ controls the clustering of physical nodes around $y=0$.
The limit $\beta\rightarrow0$ recovers the linear map
$y=y_{\max}x$. If $\mathbf{D}_x$ denotes the Chebyshev first-derivative
matrix on $[-1,1]$, the physical first-derivative matrix is constructed as
\begin{equation}
    \mathbf{D}_y
    =
    \operatorname{diag}
    \left[
        \left(
        \frac{\mathrm{d} y}{\mathrm{d} x}
        \right)^{-1}
    \right]
    \mathbf{D}_x,
    \qquad
    \frac{\mathrm{d} y}{\mathrm{d} x}
    =
    y_{\max}
    \frac{
        \beta\cosh(\beta x)
    }{
        \sinh\beta
    }.
    \label{eq:wake_mapped_D1}
\end{equation}
The higher-order nodal differentiation matrices are obtained from $\mathbf{D}^{(n)} = \mathbf{D}_y^n.$

The reduced eigenproblem is
\begin{equation}
    \left[
        \mathbf{A}_w(\alpha,Re,\lambda)
        -
        \omega\mathbf{B}_w(\alpha)
    \right]
    \mathbf{q}_w
    =
    \mathbf{0},
    \label{eq:wake_discrete}
\end{equation}
where
\begin{align}
    \mathbf{A}_w
    &=
    \alpha\mathbf{U}\mathbf{L}_{\alpha}
    -
    \alpha\mathbf{U}''
    +
    \frac{\mathrm{i}}{Re}\mathbf{M}_{\alpha}, \notag   \\
    \mathbf{B}_w
    &=
    \mathbf{L}_{\alpha}, \notag   \\
    \mathbf{L}_{\alpha}
    &=
    \mathbf{D}^{(2)}-\alpha^2\mathbf{I}, \notag   \\
    \mathbf{M}_{\alpha}
    &=
    \mathbf{D}^{(4)}
    -
    2\alpha^2\mathbf{D}^{(2)}
    +
    \alpha^4\mathbf{I}.    \notag
\end{align}
Here $\mathbf{U}$ and $\mathbf{U}''$ are diagonal matrices containing the
base velocity and its second derivative at the retained physical nodes.
The analytical wavenumber derivatives required by the adjoint saddle
condition are
\begin{align}
    \partial_{\alpha}\mathbf{L}_{\alpha}
    &=
    -2\alpha\mathbf{I},
    \\
    \partial_{\alpha}\mathbf{M}_{\alpha}
    &=
    -4\alpha\mathbf{D}^{(2)}
    +
    4\alpha^3\mathbf{I},
    \label{eq:wake_dM}
    \\
    \partial_{\alpha}\mathbf{A}_w
    &=
    \mathbf{U}\mathbf{L}_{\alpha}
    +
    \alpha\mathbf{U}
    \partial_{\alpha}\mathbf{L}_{\alpha}
    -
    \mathbf{U}''
    +
    \frac{\mathrm{i}}{Re}
    \partial_{\alpha}\mathbf{M}_{\alpha},
    \label{eq:wake_dA}
    \\
    \partial_{\alpha}\mathbf{B}_w
    &=
    -2\alpha\mathbf{I}.
    \label{eq:wake_dB}
\end{align}
Consequently,
\begin{equation}
    \frac{\mathrm{d}\omega}{\mathrm{d}\alpha}
    =
    \frac{
    (\mathbf{q}_w^{\dagger})^H
    \left(
        \partial_{\alpha}\mathbf{A}_w
        -
        \omega\partial_{\alpha}\mathbf{B}_w
    \right)
    \mathbf{q}_w
    }{
    (\mathbf{q}_w^{\dagger})^H
    \mathbf{B}_w\mathbf{q}_w
    }.
    \label{eq:wake_sensitivity}
\end{equation}
The continuation parameters are taken as
$(R_1,R_2)=(Re,\lambda)$, and the gauge-fixed boundary system in
Eq.~\eqref{eq:boundary_system_complex} is applied without any
wake-specific normalisation.

The calculations use 
$N=100$,
$y_{\max}=20$,
and
$\beta=3$,
for which the nearest non-zero physical node to the wake centre is at
$\Delta y_0= 0.1884$ and eleven full-grid nodes lie in
$|y|\le1$. Resolution, mapping-parameter, and domain-truncation studies are
reported in \ref{app:wake_convergence}. In particular, relative to
the $N=120$ calculation, the production-grid critical Reynolds number differs
by $\sim 10^{-7}$, while changing the truncated-domain half-width from
$y_{\max}=20$ to $30$ at approximately fixed wake-core spacing changes
$Re_c$ by only $ \sim 10^{-7}$ in relative terms.

At $\lambda=0.95$, the gauge-fixed neutral-saddle correction gives
$Re_c=333.2348$,
$\alpha_s=0.9342  -  0.8624\,\mathrm{i}$,
$\omega_s= 0.5260  +  1.32\times10^{-23}\,\mathrm{i}$.
The direct, full-adjoint, saddle, and biorthogonality residuals at this point
are $3.458\times10^{-12}$, $1.193\times10^{-11}$,
$1.152\times10^{-12}$, and $1.503\times10^{-12}$, respectively. This
converged state is used as the starting point for the two-sided continuation
below.

\subsection{Neutral-saddle boundary and finite-difference cross-validation}
\label{subsec:wake_continuation}

\begin{figure}[htbp]
    \centering
    \includegraphics[width=0.6\textwidth]{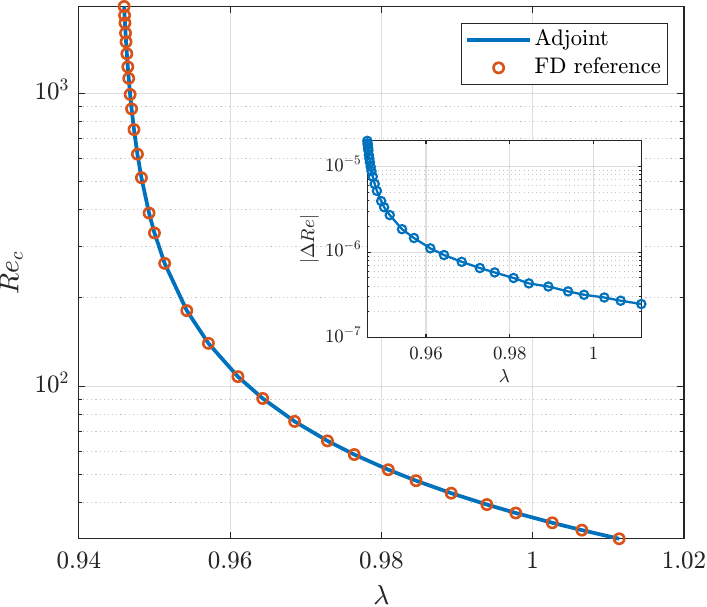}
    \caption{Gaussian-wake neutral-saddle boundary obtained by adjoint continuation. 
    Symbols show 31 finite-difference neutral-saddle
    corrections used for local cross-validation of the adjoint saddle
    formulation.
    Inset shows the absolute Reynolds number difference
$|\Delta Re|=|Re_{\mathrm{adjoint}}-Re_{\mathrm{FD}}|$
between the adjoint continuation and the finite-difference reference
calculations.
    }
    \label{fig:wake_boundary}
\end{figure}

Starting from the converged point above, the neutral-saddle equations are
continued in the $(Re,\lambda)$ plane by the scaled pseudo-arclength method
of Sec.~\ref{subsec:aici_palc}. The two orientations of the initial
nullspace tangent are followed and merged into a single connected branch.
Over the prescribed continuation window, 139 unique boundary points are
accepted, spanning
$30.1498 \le Re \le 1975.84$ and $0.945992 \le \lambda \le 1.011430$.
The two endpoints correspond to the prescribed parameter window and should
not be interpreted as terminal points of the global neutral-saddle branch.

The continuation remains numerically regular throughout the computed
segment. No pseudo-arclength step is rejected, the mean number of Newton
iterations per accepted point is 2.81, and no corrector requires more than
three iterations. The maximum residuals over the complete computed branch
are
\begin{align}
    \max\|\mathbf{L}_w\mathbf{q}_w\|_2
    &=
    1.779\times10^{-11}, \notag
    \\
    \max\|\mathbf{L}_w^H\mathbf{q}_w^\dagger\|_2
    &=
    9.617\times10^{-11}, \notag
    \\
    \max
    \left|
    (\mathbf{q}_w^\dagger)^H
    \mathbf{C}_w\mathbf{q}_w
    \right|
    &=
    4.196\times10^{-12}, \notag
    \\
    \max|\omega_i|
    &=
    1.000\times10^{-22}, \notag
    \\
    \max
    \frac{
        \|\mathbf{J}_Z\mathbf{t}^{(z)}\|_2
    }{
        \max(1,\|\mathbf{J}_Z\|_F)
    }
    &=
    4.395\times10^{-17}. \notag
\end{align}

As a local cross-validation of the adjoint saddle formulation, a separately
formulated finite-difference neutral-saddle calculation is performed at 31
values of $\lambda$ sampled along the continued branch. At each selected
point, the frequency and direct mode are retained as branch-tracking
anchors, while the neutral saddle is re-corrected using a residual that does
not contain the adjoint eigenvector or the analytical eigenvalue derivative.
At fixed $\lambda$, the reference unknown is $\mathbf{y} = (\alpha_r,\alpha_i,Re)^T$,
and the nonlinear residual is
\begin{equation}
    \mathbf{R}_{\mathrm{FD}}(\mathbf{y})
    =
    \begin{pmatrix}
        \operatorname{Re}
        \left[
        \left(
        \mathrm{d}\omega/\mathrm{d}\alpha
        \right)_{\mathrm{FD}}
        \right]
        \\
        \operatorname{Im}
        \left[
        \left(
        \mathrm{d}\omega/\mathrm{d}\alpha
        \right)_{\mathrm{FD}}
        \right]
        \\
        \operatorname{Im}(\omega)
    \end{pmatrix}
    =
    \mathbf{0}.
    \label{eq:wake_reference_residual}
\end{equation}
The finite-difference group velocity is evaluated from branch-tracked
temporal eigensolves using centred differences. Neither an adjoint
eigenvector nor the analytical matrices
$\partial_{\alpha}\mathbf{A}_w$ and
$\partial_{\alpha}\mathbf{B}_w$ enter
Eq.~\eqref{eq:wake_reference_residual}. The $3\times3$ Newton Jacobian of
the finite-difference residual is also approximated by centred differences.
The comparison therefore tests the adjoint saddle equation against a
separately formulated finite-difference residual on the same discrete wake
operator.

Figure~\ref{fig:wake_boundary} shows the continued branch together with the
31 finite-difference reference points. The agreement extends over nearly two
orders of magnitude in $Re$, rather than being restricted to a neighbourhood
of the initial neutral saddle.
For a quantity $g\in\{Re_c,\alpha_s,\omega_s\}$, the pointwise
cross-validation is summarised by
\begin{align}
    E_{\infty}(g)
    &=
    \max_{\ell}
    \left|
        g_{\ell}^{\mathrm{adj}}
        -
        g_{\ell}^{\mathrm{FD}}
    \right|,
    \label{eq:wake_boundary_Linf}
    \\
    E_{\mathrm{RMS}}(g)
    &=
    \left[
    \frac{1}{N_{\lambda}}
    \sum_{\ell=1}^{N_{\lambda}}
    \left|
        g_{\ell}^{\mathrm{adj}}
        -
        g_{\ell}^{\mathrm{FD}}
    \right|^2
    \right]^{1/2}.
    \label{eq:wake_boundary_RMS}
\end{align}
All 31 finite-difference corrections converge. The quantitative comparison
is listed in Table~\ref{tab:wake_boundary_errors}. The maximum absolute
difference in the critical Reynolds number is
$2.004\times10^{-5}$, corresponding to a maximum relative difference of
$1.046\times10^{-8}$. The maximum differences in the complex saddle
wavenumber and frequency are $2.572\times10^{-8}$ and
$4.290\times10^{-10}$, respectively. The largest residual of the
finite-difference reference solver is $8.513\times10^{-9}$. The two
formulations therefore give closely consistent neutral-saddle states over
the complete set of sampled $\lambda$ values.

\begin{table}[htbp]
    \centering
    \caption{Finite-difference cross-validation of the adjoint 
    neutral-saddle boundary at 31 sampled values of $\lambda$.}
    \label{tab:wake_boundary_errors}
    \begin{tabular}{lcc}
        \toprule
        Quantity
        &
        $E_{\infty}$
        &
        $E_{\mathrm{RMS}}$
        \\
        \midrule
        $Re_c$
        &
        $2.004\times10^{-5}$
        &
        $8.790\times10^{-6}$
        \\
        $\alpha_s$
        &
        $2.572\times10^{-8}$
        &
        $1.075\times10^{-8}$
        \\
        $\omega_s$
        &
        $4.290\times10^{-10}$
        &
        $2.140\times10^{-10}$
        \\
        \bottomrule
    \end{tabular}
\end{table}

Although pseudo-arclength continuation is designed to pass parameter folds,
no actual $Re$- or $\lambda$-turning point is crossed on the computed wake
segment. The consistently oriented physical parameter-plane tangent
$\mathbf{t}^{(p)}$ preserves the signs of both parameter components
throughout the calculation, with no zero crossing of either
$t_{Re}^{(p)}$ or $t_{\lambda}^{(p)}$. Toward the high-$Re$ end,
$t_{\lambda}^{(p)}$ approaches zero while $t_{Re}^{(p)}$ remains non-zero,
showing that the projection of the solution manifold onto the
$(Re,\lambda)$ plane becomes nearly vertical. Thus, $\lambda$ becomes a
poorly conditioned natural continuation parameter even though a true fold
is not encountered in the present window. The pseudo-arclength
parameterisation remains regular and does not require a manual switch of
physical continuation parameter.

\subsection{Computational efficiency relative to Briggs nested saddle scanning}
\label{subsec:wake_efficiency}

The finite-difference calculations in the preceding subsection are intended
as local cross-validation of selected neutral-saddle states and do not
constitute an independent boundary-construction procedure. To assess the
cost of constructing the complete boundary, the direct adjoint trace is
compared with a nested complex-wavenumber and parameter-plane
saddle-scanning procedure.

The two approaches use the same mapped Orr--Sommerfeld discretisation,
Eq.~\eqref{eq:wake_discrete}, 
and target the saddle family identified from
the neutral point at $\lambda=0.95$. No points from the computed
adjoint  boundary are supplied to the scanning calculation. At each
sampled parameter pair $(Re,\lambda)$, a temporal eigenvalue sheet is first
followed on a fixed $11\times11$ complex-$\alpha$ grid covering
$0.25\le\alpha_r\le1.45,$ and $-1.55\le\alpha_i\le-0.15.$
Candidate stationary points are detected from small values of the
finite-difference group velocity
$g_{\mathrm{FD}}(\alpha) = \frac{\omega(\alpha+h_\alpha) - \omega(\alpha-h_\alpha)
    }{2h_\alpha },$
and are locally refined by solving
$\operatorname{Re}g_{\mathrm{FD}}
=
\operatorname{Im}g_{\mathrm{FD}}
=0$
with a finite-difference Newton iteration in
$(\alpha_r,\alpha_i)$. Continuity in the saddle coordinates, eigenvalue and
modal overlap is used to retain the selected saddle family. The absolute
growth rate
\begin{equation}
    \gamma(Re,\lambda)
    =
    \operatorname{Im}\omega_s(Re,\lambda)
\end{equation}
is then evaluated throughout a two-dimensional parameter grid, and the
neutral boundary is reconstructed a posteriori from the zero contour
$\gamma=0$. The comparison therefore represents a nested
scanning-and-reconstruction workflow rather than an optimally warm-started
saddle-continuation algorithm.

All timing measurements are performed in MATLAB R2020a on a system equipped
with an AMD Ryzen 7 PRO 4750U processor and 16~GB of system memory. The
adjoint and scanning calculations are executed on the same hardware
using the same spatial discretisation. 
For the adjoint calculation, the timed region begins after the initial fixed-$\lambda$
neutral-saddle correction and includes the two directional
pseudo-arclength traces. For the scanning calculations, the timed region
contains the complex-$\alpha$ searches, local saddle refinements over the
parameter grid and reconstruction of the neutral zero contour. 

\begin{figure}[htbp]
    \centering
    \includegraphics[width=0.6\textwidth]
    {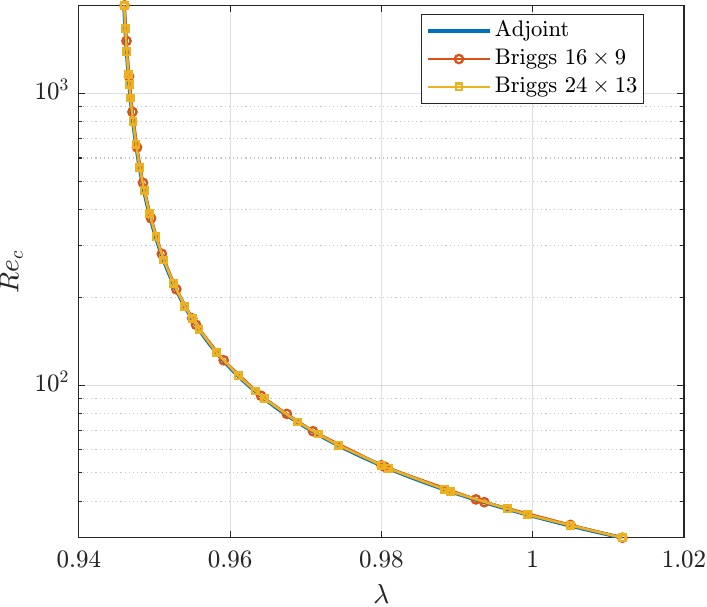}
    \caption{Gaussian-wake neutral-saddle boundary obtained by direct
    adjoint continuation and by nested saddle scanning on
    $16\times9$ and $24\times13$ $(Re,\lambda)$ grids. Both scanning
    calculations use the same $11\times11$ complex-$\alpha$ search grid.
    The curves are shown over the common Reynolds-number interval used for
    the geometric error measures.}
    \label{fig:wake_briggs_palc_comparison}
\end{figure}

Six parameter-plane resolutions are considered,
$N_{Re} \times N_{\lambda} = $
$12\times7,  16\times9, 20\times11, 24\times13, 28\times15, 32\times17,$
where $N_{Re}$ and $N_{\lambda}$ denote the numbers of $Re$ and
$\lambda$ nodes, respectively. The $Re$ nodes are logarithmically
distributed over $30\le Re\le2000$, whereas the $\lambda$ nodes are uniformly
distributed over $0.93\le\lambda\le1.03$. The complex-$\alpha$ grid and
local saddle-refinement settings are held fixed throughout. Every sampled
parameter point converges successfully and each resolution produces one
connected neutral contour.
Figure~\ref{fig:wake_briggs_palc_comparison} compares two representative
scanning reconstructions with the direct adjoint boundary. The
$16\times9$ scan already recovers the global geometry of the neutral
boundary, while refinement to $24\times13$ brings the reconstructed contour
closer to the direct continuation result.

The reconstruction error is measured geometrically in the normalised
$(\lambda,\log Re)$ plane,
\begin{equation}
    \xi_\lambda
    =
    \frac{\lambda-\lambda_{\min}}
    {\lambda_{\max}-\lambda_{\min}},
    \qquad
    \xi_R
    =
    \frac{\log Re-\log Re_{\min}}
    {\log Re_{\max}-\log Re_{\min}},
    \label{eq:wake_boundary_normalisation}
\end{equation}
where the normalisation limits are the fixed parameter-scan bounds. Let
$\mathcal{B}_{P}$ and $\mathcal{B}_{B}$ denote the adjoint and scanning
boundaries, respectively. The symmetric Hausdorff-type distance is
\begin{equation}
    E_H
    =
    \max
    \left\{
        \sup_{\mathbf{x}\in\mathcal{B}_{P}}
        d(\mathbf{x},\mathcal{B}_{B}),
        \,
        \sup_{\mathbf{y}\in\mathcal{B}_{B}}
        d(\mathbf{y},\mathcal{B}_{P})
    \right\},
    \label{eq:wake_boundary_hausdorff}
\end{equation}
where $d(\mathbf{x},\mathcal{B})$ is the minimum Euclidean distance from
$\mathbf{x}$ to the piecewise-linear boundary $\mathcal{B}$. A symmetric
geometric RMS distance, $E_{\mathrm{RMS}}^{(g)}$, is evaluated from the two
directed point-to-boundary distance sets. The curves are resampled with a
maximum normalised spacing of $10^{-4}$, and the comparison is restricted to
the common interval
$30.1498\le Re\le1975.84$ so that the error is not controlled by differences
in the discrete termination of the two algorithms.

The cost and boundary errors are summarised in
Table~\ref{tab:wake_efficiency}. The complete two-sided adjoint trace
requires $28.33$~s for 139 accepted boundary points. In contrast, the
coarsest $12\times7$ scan already requires $229.34$~s, or $8.10$ times the
adjoint wall time, while giving
$E_H=4.073\times10^{-3}$. Increasing the parameter-grid resolution reduces
the geometric error monotonically, but the cost increases rapidly because
the complex-wavenumber search and saddle refinement are repeated at every
additional parameter point. On the finest $32\times17$ grid,
$E_H$ is reduced to $5.362\times10^{-4}$, whereas the wall time rises to
$1479.67$~s, corresponding to $52.23$ times the adjoint cost.

\begin{table}[htbp]
    \centering
    \caption{Cost and geometric error of the nested saddle scans. The
    adjoint  boundary requires $28.33$~s and is used as the geometric
    reference.}
    \label{tab:wake_efficiency}
    \setlength{\tabcolsep}{3.5pt}
    \begin{tabular}{@{}cccccc@{}}
        \toprule
        Grid
        &
        $N_{\rm p}$
        &
        $T$ (s)
        &
        $T/T_{\rm adjoint}$
        &
        $E_H$
        &
        $E_{\mathrm{RMS}}^{(g)}$
        \\
        \midrule
        $12\times7$
        &
        84
        &
        229.34
        &
        8.10
        &
        $4.073\times10^{-3}$
        &
        $2.193\times10^{-3}$
        \\
        $16\times9$
        &
        144
        &
        397.88
        &
        14.04
        &
        $2.332\times10^{-3}$
        &
        $1.275\times10^{-3}$
        \\
        $20\times11$
        &
        220
        &
        601.93
        &
        21.25
        &
        $1.419\times10^{-3}$
        &
        $7.622\times10^{-4}$
        \\
        $24\times13$
        &
        312
        &
        866.25
        &
        30.58
        &
        $9.937\times10^{-4}$
        &
        $5.122\times10^{-4}$
        \\
        $28\times15$
        &
        420
        &
        1144.00
        &
        40.38
        &
        $7.281\times10^{-4}$
        &
        $3.833\times10^{-4}$
        \\
        $32\times17$
        &
        544
        &
        1479.67
        &
        52.23
        &
        $5.362\times10^{-4}$
        &
        $2.830\times10^{-4}$
        \\
        \bottomrule
    \end{tabular}
\end{table}

\begin{figure}[htbp]
    \centering
    \includegraphics[width=1\textwidth]
    {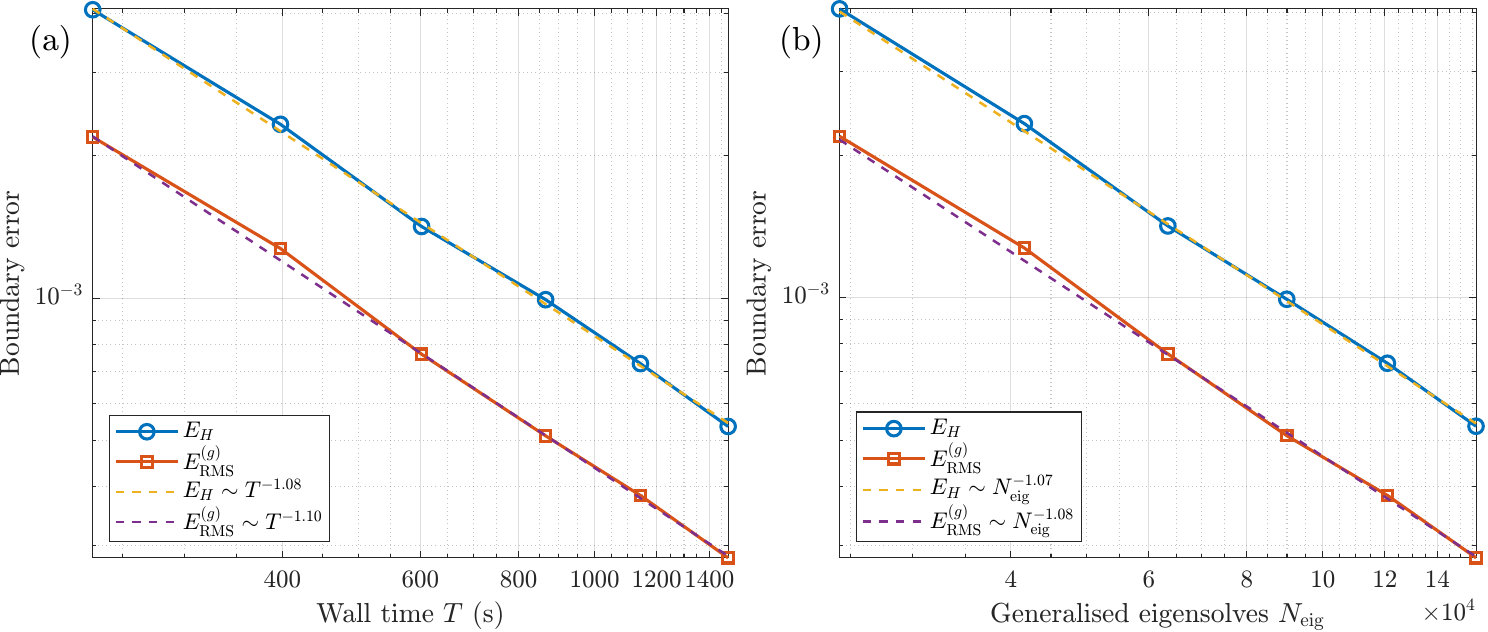}
    \caption{
    (a) Boundary error of the nested Gaussian-wake saddle scans as a
    function of wall time and (b) the total number of generalised
    eigensolves. The $(Re,\lambda)$ grid is refined while the
    complex-$\alpha$ search grid and local saddle-refinement settings are
    held fixed. Dashed lines show power-law fits over the finest four
    parameter grids.}
    \label{fig:wake_briggs_cost_accuracy}
\end{figure}

Figure~\ref{fig:wake_briggs_cost_accuracy} shows the resulting
cost--accuracy relation. Over the tested parameter-grid range, increasing the
scan cost gives an approximately algebraic reduction of the reconstructed
boundary error. Fits over the finest four grids give
\begin{equation}
    E_H\propto T^{-1.08},
    \qquad
    E_{\mathrm{RMS}}^{(g)}\propto T^{-1.10},
    \label{eq:wake_time_accuracy_scaling}
\end{equation}
while the corresponding eigensolve-count fits are
$E_H\propto N_{\mathrm{eig}}^{-1.07}$ and
$E_{\mathrm{RMS}}^{(g)}\propto N_{\mathrm{eig}}^{-1.08}$.
Here, $N_{\mathrm{eig}}$ denotes the total number of generalised eigenvalue solves performed during the nested complex-wavenumber search and local saddle-refinement procedure.
These relations are empirical trends over the resolutions considered here,
rather than formal asymptotic convergence orders. 
Extrapolating the fitted $E_H$--$T$ relation for the finest four scans to a
target boundary error of $E_H=10^{-6}$ gives an estimated nested-scanning
time of approximately $5.0\times10^{5}$~s. Relative to the $28.33$~s
required by the direct adjoint trace, this corresponds to an estimated cost
ratio of approximately $1.8\times10^{4}$. 
The total scanning time is nearly proportional to the number of parameter points; equivalently, the
calculations require approximately $288$ generalised eigensolves per sampled
$(Re,\lambda)$ pair.
More specifically, the measured wall time scales with the total number of
parameter-plane nodes as
\begin{equation}
    T \propto N_{\mathrm{p}},
    \qquad
    N_{\mathrm{p}}=N_{Re}N_{\lambda},
    \label{eq:wake_grid_time_scaling}
\end{equation}
showing an approximately linear increase in computational cost with
parameter-grid size. For the simultaneous $(Re,\lambda)$ refinement used
here, this corresponds over the tested grid range to
\begin{equation}
    T \propto h_{\mathrm{eff}}^{-1.87},
    \qquad
    h_{\mathrm{eff}}
    =
    \left[
        \frac{1}{(N_{Re}-1)(N_{\lambda}-1)}
    \right]^{1/2}.
    \label{eq:wake_mesh_time_scaling}
\end{equation}
Thus, improving the scanning resolution requires increasing the number of
two-dimensional parameter samples, with the measured cost growing almost
directly with the resulting grid size.

The comparison highlights the different numerical structures of the two
approaches. The nested procedure first determines a saddle at every sampled
point in the two-dimensional $(Re,\lambda)$ plane and only then reconstructs
the neutral contour from the field
$\operatorname{Im}\omega_s(Re,\lambda)$. Improving the boundary accuracy
therefore requires additional parameter-plane sampling and repeated local
spectral calculations away from the boundary itself. By contrast, the
adjoint residual embeds the stationary condition directly in the augmented
system and adjoint method advances only along the one-dimensional neutral-saddle
manifold. The computational advantage of the present formulation therefore
arises not merely from a different evaluation of
$\frac{\mathrm{d}\omega}{\mathrm{d}\alpha}$, but from replacing two-dimensional
scan-and-reconstruction by direct continuation of the boundary itself.

\subsection{Spatial pinch topology}
\label{subsec:wake_pinch}

\begin{figure}[htbp]
    \centering
    \includegraphics[width=0.9\textwidth]{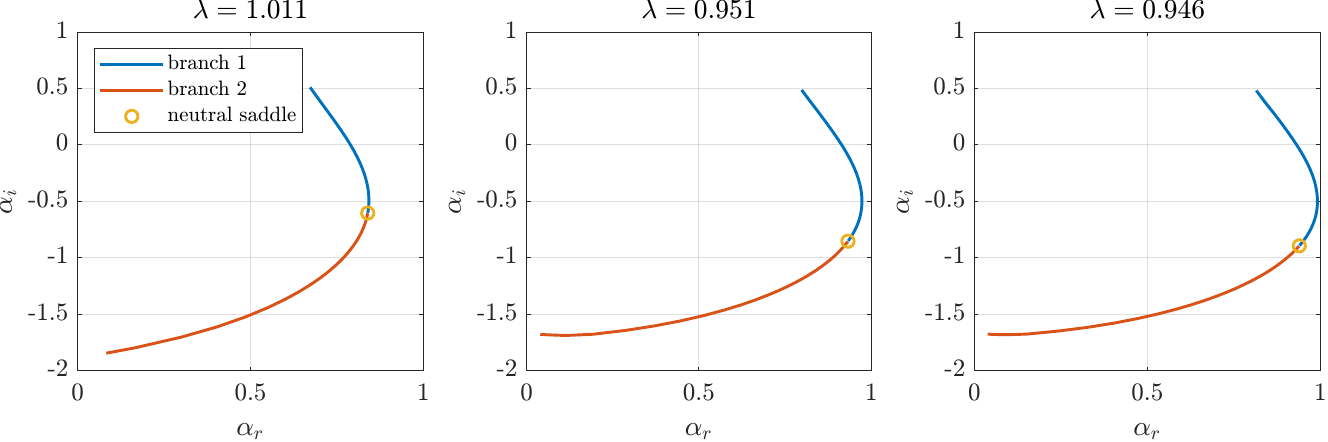}
    \caption{Tracked spatial-root trajectories for three representative
    points $ (Re,\lambda)=(30.15,1.011), (262.47,0.951), (1975.84,0.946)$ on the Gaussian-wake neutral-saddle branch. The complex frequency
    is displaced according to
    $\omega=\omega_s+\mathrm{i}\gamma$ and the two branches are followed toward the
    neutral saddle as $\gamma\rightarrow0^{+}$. In each case the branches
    originate from opposite halves of the complex $\alpha$ plane and
    coalesce with the square-root scaling expected for a simple double root.}
    \label{fig:wake_pinch}
\end{figure}

The extended system identifies neutral stationary points of the temporal
dispersion relation, but the AI/CI interpretation additionally requires the
relevant spatial branches to form a Briggs--Bers pinch
point~\cite{huerre1985absolute,huerre1990local}. Representative spatial-root
trajectories are therefore examined at the first, middle, and last points of
the continued wake segment.

For fixed complex $\omega$, the reduced Orr--Sommerfeld operator is a
fourth-degree matrix polynomial in $\alpha$,
\begin{equation}
    \mathbf{P}
    (\alpha;\omega,Re,\lambda)
    =
    \sum_{j=0}^{4}
    \alpha^j\mathbf{P}_j,
    \label{eq:wake_spatial_polynomial}
\end{equation}
with
\begin{align}
    \mathbf{P}_4 &= \frac{\mathrm{i}}{Re}\mathbf{I}, \qquad \qquad\quad
    \mathbf{P}_3 = -\mathbf{U},\qquad 
    \mathbf{P}_2 = -\frac{2\mathrm{i}}{Re}\mathbf{D}^{(2)} + \omega\mathbf{I}, \notag  \\
    \mathbf{P}_1 &= \mathbf{U}\mathbf{D}^{(2)} - \mathbf{U}'', \qquad 
    \mathbf{P}_0 = \frac{\mathrm{i}}{Re}\mathbf{D}^{(4)} - \omega\mathbf{D}^{(2)}.\notag 
\end{align}
A first companion linearisation converts
Eq.~\eqref{eq:wake_spatial_polynomial} into a generalised eigenvalue problem for the spatial roots
$\alpha$. 
For each selected neutral saddle, the frequency is displaced to
$\omega = \omega_s+\mathrm{i}\gamma$, $10^{-6}\le\gamma\le0.4$,
and the two roots closest to $\alpha_s$ at the smallest $\gamma$ are
tracked continuously as $\gamma$ increased.
The three test points are
$ (Re,\lambda)=(30.15,1.011), (262.47,0.951), (1975.84,0.946).$
As illustrated in Fig.  \ref{fig:wake_pinch}, at $\gamma=0.4$, the two tracked roots lie in opposite halves of the complex $\alpha$ plane for all three cases. 
As $\gamma\rightarrow0^{+}$, the two branches approach the computed
neutral saddle. Near a simple double root, the local expansion
\begin{equation}
    D(\alpha,\omega)
    \simeq
    D_{\omega}
    (\omega-\omega_s)
    +
    \frac{1}{2}
    D_{\alpha\alpha}
    (\alpha-\alpha_s)^2
\end{equation}
predicts
\begin{equation}
    |\alpha_1-\alpha_2|
    \propto
    |\omega-\omega_s|^{1/2}.
    \label{eq:wake_square_root_splitting}
\end{equation}
A log--log fit to the first 15 low-$\gamma$ points all give exponents
$0.499$ for the three cases. 
The opposite-half-plane origin of the two branches and their
square-root coalescence at $\alpha_s$ provide independent spatial-root
evidence that the representative stationary points are the relevant pinch
points rather than unrelated algebraic saddles.

In this section, the Gaussian-wake results assess the proposed method in a
matrix-valued, non-normal eigenvalue problem. The sinh-mapped collocation
exhibits convergence with respect to resolution, mapping, and domain
truncation, while the complete continued segment agrees with the
finite-difference neutral-saddle cross-validation to approximately $10^{-8}$
in relative critical Reynolds number. Finally, representative spatial-root
calculations recover the pinch topology required for the physical AI/CI
interpretation. The wake problem therefore provides a matrix-level
continuation and cross validation test before the method is applied to the
coupled viscoelastic-film eigenproblem.

\section{Application to the spatio-temporal instability problem of viscoelastic liquid films}
\label{sec:film}

The preceding Ginzburg--Landau and Gaussian-wake problems test the
adjoint saddle formulation in an analytical scalar limit and in a
non-normal matrix eigenvalue problem, respectively. We now apply the same
gauge-fixed direct--adjoint continuation system to a coupled viscoelastic
free-surface problem. The purpose of this example is twofold. First, it
demonstrates the frequency-affine formulation for a multi-field state
containing hydrodynamic, constitutive, and interfacial variables. Second, it
provides a genuine turning-point test in which the neutral-saddle manifold
contains folds of its Reynolds-number projection.

\subsection{Flow model, nondimensionalisation, and base state}
\label{subsec:film_model}

We consider a two-dimensional incompressible Oldroyd--B liquid film flowing
on the underside of a plane inclined by an angle $\theta$ to the horizontal.
The dimensional mean film thickness is $h_0$. The streamwise and wall-normal
coordinates are $x^*$ and $y^*$, respectively, and the deformed interface is
$y^*=h^*(x^*,t^*)$.

The dimensional conservation equations are
\begin{align}
    \partial_i u_i^* &=0,
    \label{eq:film_dim_cont}\\
    \rho\left(
    \partial_t u_i^*
    +
    u_j^*\partial_j u_i^*
    \right)
    &=
    -\partial_i p^*
    +
    \partial_j\tau_{ij}^*
    +
    \rho g_i,
    \label{eq:film_dim_mom}
\end{align}
and the Oldroyd--B constitutive equation is
\begin{equation}
    \boldsymbol{\tau}^*
    +
    \lambda_1\overset{\nabla}{\boldsymbol{\tau}^*}
    =
    2\mu
    \left(
    \mathbf{e}^*
    +
    \lambda_2\overset{\nabla}{\mathbf{e}^*}
    \right),
    \label{eq:film_oldroyd_dim}
\end{equation}
where
$\mathbf{e}^*=[\nabla\mathbf{u}^*+(\nabla\mathbf{u}^*)^T]/2$,
$\lambda_1$ is the relaxation time,
$\lambda_2$ is the retardation time, and
$\mu=\mu_s+\mu_p$. 
The ratio $S=\frac{\lambda_2}{\lambda_1} = \frac{\mu_s}{\mu_s+\mu_p}$
measures the solvent fraction.

At $y^*=0$, no slip and no penetration are imposed. At
$y^*=h^*(x^*,t^*)$, the kinematic condition and the tangential and normal
stress balances are
\begin{align}
    \partial_t h^*
    +
    u^*\partial_{x^*}h^*
    &=
    v^*,
    \label{eq:film_dim_kinematic}\\
    \tau_{ij}^*n_it_j
    &=
    0,
    \label{eq:film_dim_tangential}\\
    p_a-p^*+\tau_{ij}^*n_in_j
    &=
    \sigma
    \frac{\partial_{x^*x^*}h^*}
    {\left[1+(\partial_{x^*}h^*)^2\right]^{3/2}}.
    \label{eq:film_dim_normal}
\end{align}

We use $h_0$ and the Nusselt surface velocity $u_0 = \frac{\rho g\sin\theta\,h_0^2}{2\mu}$
as length and velocity scales. Pressure is scaled by $\rho u_0^2$ and
extra stress by $\frac{\mu u_0}{h_0}$. The dimensionless groups are
\begin{equation}
    Re=\frac{\rho u_0h_0}{\mu},
    \qquad
    De=\frac{\lambda_1u_0}{h_0},
    \qquad
    We=\frac{2\sigma}{\rho gh_0^2\sin\theta}.
    \label{eq:film_dimensionless_groups}
\end{equation}
The dimensionless momentum equations are
\begin{align}
    \partial_xu+\partial_yv&=0,
    \label{eq:film_cont}\\
    Re\left(
    \partial_tu+u\partial_xu+v\partial_yu
    \right)
    &=
    -Re\,\partial_xp
    +
    \partial_x\tau_{xx}
    +
    \partial_y\tau_{xy}
    +
    2,
    \label{eq:film_mom_x}\\
    Re\left(
    \partial_tv+u\partial_xv+v\partial_yv
    \right)
    &=
    -Re\,\partial_yp
    +
    \partial_x\tau_{xy}
    +
    \partial_y\tau_{yy}
    +
    2\cot\theta.
    \label{eq:film_mom_y}
\end{align}
For the Oldroyd-B fluid, the dimensionless constitutive equations for the stress components are
\begin{align}
  \begin{split}
    \tau_{xx} + De\bigl[\partial_t\tau_{xx} &+ (u\partial_x + v\partial_y)\tau_{xx} - 2(\tau_{xx}\partial_x u + \tau_{xy}\partial_y u)\bigr] \\
    &= 2\partial_x u + 2De S\bigl[\partial_{xt} u + u\partial_{xx}u + v\partial_{xy}u \\
    &\qquad - (\partial_y u + \partial_x v)\partial_y u - 2(\partial_x u)^2\bigr],
  \end{split}
  \label{eq:tau_xx_full}
  \\[1ex]
  \begin{split}
    \tau_{xy} + De\bigl[\partial_t\tau_{xy} &+ (u\partial_x + v\partial_y)\tau_{xy} - \tau_{xx}\partial_x v - \tau_{yy}\partial_y u\bigr] \\
    &= (\partial_y u + \partial_x v) + De S\bigl[\partial_{yt}u + \partial_{xt}v + u\partial_{xy}u + u\partial_{xx}v \\
    &\qquad + v\partial_{yy}u + v\partial_{xy}v - 2\partial_x u\partial_x v - 2\partial_y u\partial_y v\bigr],
  \end{split}
  \label{eq:tau_xy_full}
  \\[1ex]
  \begin{split}
    \tau_{yy} + De\bigl[\partial_t\tau_{yy} &+ (u\partial_x + v\partial_y)\tau_{yy} - 2(\tau_{xy}\partial_x v + \tau_{yy}\partial_y v)\bigr] \\
    &= 2\partial_y v + 2De S\bigl[\partial_{yt} v + u\partial_{xy}v + v\partial_{yy}v \\
    &\qquad - (\partial_y u + \partial_x v)\partial_x v - 2(\partial_y v)^2\bigr],
  \end{split}
  \label{eq:tau_yy_full}
\end{align}

The Nusselt base state is
\begin{align}
    \bar{u}(y)&=-y^2+2y,
    &
    \bar{v}(y)&=0,
    \label{eq:film_base_velocity}\\
    \bar{p}(y)&=\frac{2\cot\theta}{Re}(y-1),
    \label{eq:film_base_pressure}\\
    \bar{\tau}_{xx}(y)
    &=
    8De(1-S)(1-y)^2,
    &
    \bar{\tau}_{xy}(y)
    &=
    2(1-y),
    &
    \bar{\tau}_{yy}(y)&=0.
    \label{eq:film_base_stress}
\end{align}
The velocity profile is independent of elasticity, whereas the base
streamwise normal stress varies explicitly with $De$ and $S$.

\subsection{Frequency-affine multi-field eigenproblem}
\label{subsec:film_linear}

We perform a linear stability analysis by superimposing infinitesimal disturbances onto the base flow.  
The flow variables are decomposed as
\begin{align}
u &= \bar{u}(y) + u'(x,y,t), \quad v = v'(x,y,t), \quad p = \bar{p}(y) + p'(x,y,t), \nonumber\\
\tau_{ij} &= \bar{\tau}_{ij}(y) + \tau'_{ij}(x,y,t), \quad h = 1 + \eta(x,t).
\end{align} 
The disturbances are assumed to be wave‑like in the streamwise direction,
\begin{equation}
(u',v',p',\tau'_{ij},\eta) = \bigl(\hat u(y), \hat v(y), \hat p(y), T_{ij}(y), \hat\eta\bigr)\, e^{\mathrm{i}(k x - \omega t)}.
\end{equation}
For the spatio-temporal problem, both $k$ and $\omega$ are complex.
The linearised continuity equation $\mathrm{i}k \hat u + D\hat v = 0$ (with $D \equiv \frac{\mathrm{d}}{\mathrm{d}y}$) allows us to introduce a stream function $\phi(y)$ such that
$\hat u = D\phi$ and $\hat v = -\mathrm{i}k\phi$.
Eliminating the pressure from the linearised momentum equations by cross‑differentiation yields the Orr–Sommerfeld equation for the viscoelastic film
\begin{equation}
    \mathrm{i} Re
    \left[
    (k\bar{u}-\omega)(D^2-k^2)+2k
    \right]\phi
    =
    \mathrm{i} kDT_{xx}
    +
    (D^2+k^2)T_{xy}
    -
    \mathrm{i} kDT_{yy}.
    \label{eq:film_OS}
\end{equation}

The linearised Oldroyd--B stress equations are
\begin{align}
\mathcal{L}_{De}T_{xx}
&=
2\mathrm{i} kD\phi
+
De\left[
\mathrm{i} k\phi D\bar{\tau}_{xx}
+
2\mathrm{i} k\bar{\tau}_{xx}D\phi
+
2\bar{\tau}_{xy}D^2\phi
+
2D\bar{u}\,T_{xy}
\right]
\nonumber\\
&\quad
-
2DeS\left[
k(k\bar{u}-\omega)D\phi
+
2D\bar{u}D^2\phi
+
k^2\phi D\bar{u}
\right],
\label{eq:film_Txx}\\
\mathcal{L}_{De}T_{xy}
&=
(D^2+k^2)\phi
\left[
1+\mathrm{i} DeS(k\bar{u}-\omega)
\right]
+
De\left[
\mathrm{i} k\phi D\bar{\tau}_{xy}
+
k^2\bar{\tau}_{xx}\phi
+
D\bar{u}\,T_{yy}
\right]
\nonumber\\
&\quad
+
2\mathrm{i} kDeS
\left(
\phi+D\bar{u}D\phi
\right),
\label{eq:film_Txy}\\
\mathcal{L}_{De}T_{yy}
&=
-2\mathrm{i} kD\phi
+
2DeS\,k(k\bar{u}-\omega)D\phi
+
2De(1-S)k^2\phi D\bar{u}.
\label{eq:film_Tyy}
\end{align}
where $\mathcal{L}_{De} = 1+\mathrm{i} De(k\bar{u}-\omega).$

The wall conditions are $\phi=D\phi=0$ at $y=0$.
The interface amplitude $\hat{\eta}$ is retained as an independent unknown.
The linearised free-surface conditions at $y=1$ are
\begin{align}
    k\phi+(k-\omega)\hat{\eta}
    &=
    0,
    \label{eq:film_kinematic_bc}\\
    T_{xy}-2\hat{\eta}
    &=
    0,
    \label{eq:film_tangential_bc}\\
    \frac{Re}{k}(k-\omega)D\phi
    -
    T_{xx}
    +
    \frac{\mathrm{i}}{k}DT_{xy}
    +
    T_{yy}
    +
    \left(
    We\,k^2-2\cot\theta
    \right)\hat{\eta}
    &=
    0.
    \label{eq:film_normal_bc}
\end{align}
Eliminating $\hat{\eta}$ with the kinematic condition would introduce
$(k-\omega)^{-1}$ into the remaining boundary conditions. Retaining
$\hat{\eta}$ therefore preserves the frequency-affine form required by the
adjoint formulation.

The interval $y\in[0,1]$ is discretised with $N+1$
Chebyshev--Gauss--Lobatto points. The state is ordered as
\begin{equation}
    \mathbf{q}_{f}
    =
    \left(
        \boldsymbol{\phi}^{T},
        \mathbf{T}_{xx}^{T},
        \mathbf{T}_{xy}^{T},
        \mathbf{T}_{yy}^{T},
        \hat{\eta}
    \right)^T
    \in\mathbb{C}^{n_f},
    \qquad
    n_f=4(N+1)+1.
    \label{eq:film_state_vector}
\end{equation}
The wall and free-surface conditions replace the corresponding collocation
rows. The resulting matrix problem is
\begin{equation}
    \mathbf{L}_{f}
    (k,\omega;Re,De,S,We,\theta)
    \mathbf{q}_{f}
    =
    \mathbf{0},
    \label{eq:film_matrix_operator}
\end{equation}
with 
\begin{equation}
    \mathbf{L}_{f}
    =
    \mathbf{A}_{f}(k;Re,De,S,We,\theta)
    -
    \omega
    \mathbf{B}_{f}(k;Re,De,S,We,\theta).
    \label{eq:film_AB_form}
\end{equation}

\begin{figure}[htbp]
    \centering
    \includegraphics[width=1\textwidth]{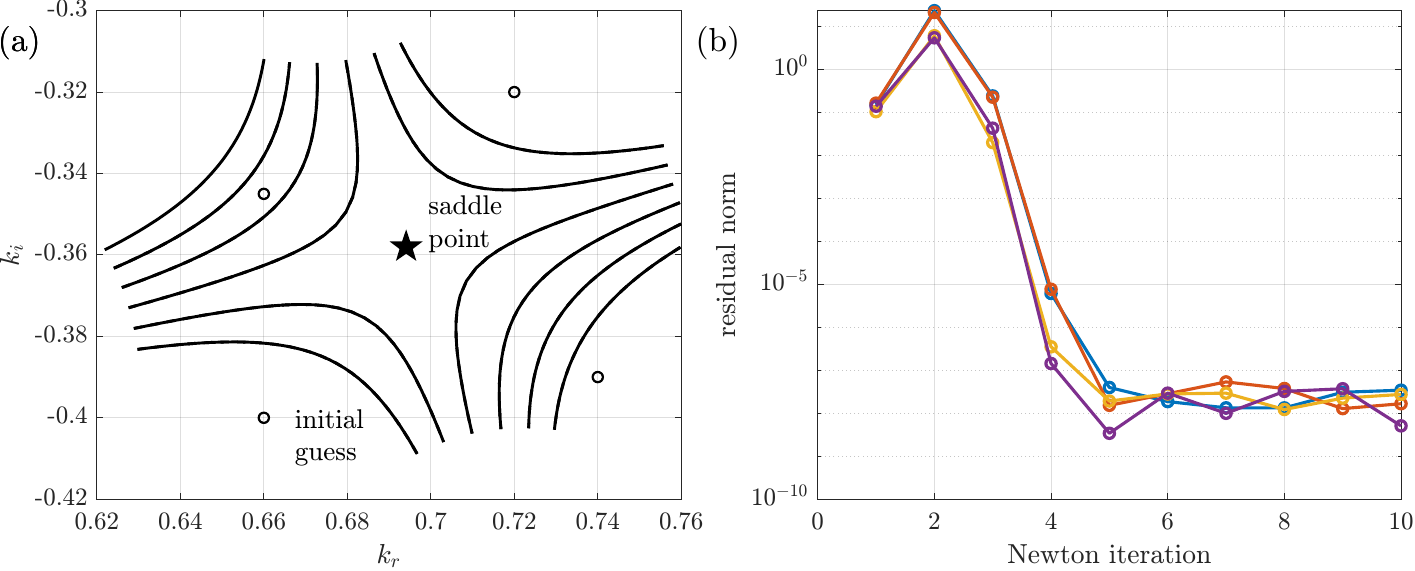}
    \caption{Representative Briggs-type spatial scan for the Oldroyd--B
    film at $Re=10$, $\theta=\frac{\pi}{16}$, $De=0$ and $We=10$.
    (a) The interacting complex-wavenumber branches approach the direct--adjoint stationary point. 
    (b) The same saddle is recovered from  nearby complex-$k$ initial guesses, demonstrating local robustness of the
    nonlinear saddle correction.}
    \label{fig:film_briggs_scan}
\end{figure}

Before tracing the AI/CI boundaries with the pseudo-arclength algorithm, we first verify the robustness of the spatial saddle point computation using Eqs.~(\ref{eq:fixed_saddle_system}) and~(\ref{eq:fixed_saddle_newton}).
The algebraic saddle condition must remain associated with the relevant
spatial pinch. 
A representative film saddle $(k_r, k_i)=(0.69409, -0.35801)$ is independently examined by a
Briggs-type scan \cite{Schmid2001} in the complex-$k$ plane, see Fig.  \ref{fig:film_briggs_scan}(a). 
The interacting spatial branches coalesce at the direct--adjoint saddle location, providing an independent
check that the selected stationary point is the physical pinch rather than an
unrelated algebraic saddle. In addition, the direct saddle solve is
restarted from multiple complex-wavenumber initial guesses in a neighbourhood
of the saddle. The perturbed calculations converged reproducibly to the same
stationary point, typically within five Newton iterations, as shown in Fig.  \ref{fig:film_briggs_scan}(b). 
The Briggs calculation is used here as a representative topology check,
rather than as a pointwise reconstruction of every continued state.

\subsection{Neutral-saddle boundaries in the $(Re,De)$ plane}
\label{subsec:film_boundary_results}

For the AI/CI boundary, $S$, $We$, and $\theta$ are fixed and the continuation
parameters are $(R_1,R_2)=(Re,De).$
The boundary residual in Eq.~\eqref{eq:boundary_system_complex} and the
scaled pseudo-arclength algorithm in Algorithm~\ref{alg:adjoint_palc} are
then applied without a film-specific continuation reformulation.
Based on convergence studies,
 the resolution of \(N=30\) and \(\frac{h_k}{1+|k|}=3.16\times10^{-4}\) is adopted. 
At \(S=0.5\), \(We=10\), \(\theta=\frac{\pi}{16}\), \(De=0\), the neutral saddle is
$Re_c = 16.7469$,
$k_s = 0.7222 - 0.4268\,\mathrm{i}$,
$\omega_s = 0.6839$
with direct, full-adjoint, saddle, and biorthogonality residuals 
\(2.38\times10^{-11}\), \(5.51\times10^{-11}\), \(4.18\times10^{-11}\), and \(1.53\times10^{-13}\), respectively. 

\begin{figure}[htbp]
    \centering
    \includegraphics[width=1\textwidth]{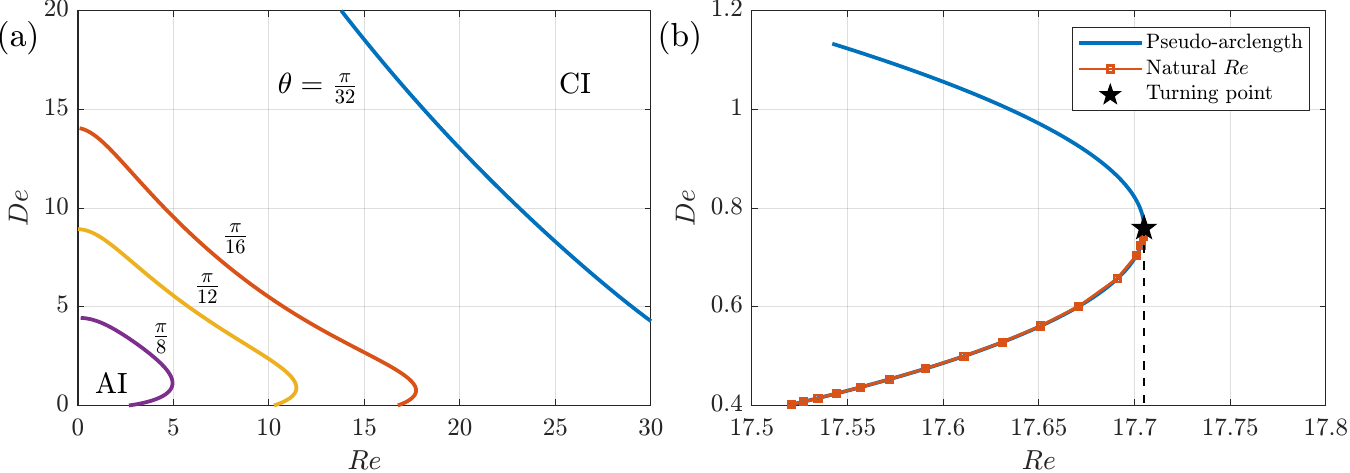}
    \caption{
    (a) Neutral-saddle boundary in the $(Re,De)$ plane for four inclination angles at $S=0.5$ and $We=10$.
    Off-boundary stationary-saddle calculations identify the smaller-$Re$ side of each local boundary
crossing as absolutely unstable, $\omega_i>0$; the larger-$Re$ side is convectively unstable.
    (b) Turning point test for the $\theta=\frac{\pi}{16}$ of Oldroyd--B film.
    Symbols mark the detected $Re$-folds. }
    \label{fig:film_boundary_four_theta}
\end{figure}

The neutral-saddle manifold is traced at
$S=0.5$, $We=10$, $\theta = \frac{\pi}{32}, \frac{\pi}{16}, \frac{\pi}{12}, \frac{\pi}{8}$.
For each inclination, the $De=0$ neutral saddle is corrected first and the
two orientations of the scaled nullspace tangent are then followed and
merged. 
Figure~\ref{fig:film_boundary_four_theta}(a) shows that all four curves develop
a single turning point in their Reynolds-number projection. The consistently
oriented parameter tangent satisfies
$t_{Re}^{(p)}: +\rightarrow0\rightarrow-$
and
$t_{De}^{(p)}>0,$
so these are $Re$-folds rather than $De$-folds. The fold coordinates and
associated neutral saddles are summarised in
Table~\ref{tab:film_fold_summary}.
The neutral-saddle equations impose $\omega_i=0$ but do not by themselves
label the two sides of the boundary. 
Stationary-saddle calculations are therefore repeated at parameter pairs displaced from the neutral curve. 
With the normal-mode convention of Eq.~\eqref{eq:normal_mode_convention}, the
smaller-$Re$ side is found to satisfy $\omega_i>0,$
and is therefore absolutely unstable. 
The opposite side is convectively unstable. This orientation is labeled in
Fig.~\ref{fig:film_boundary_four_theta}(a) by AI and CI, respectively.

\begin{table}[htbp]
    \centering
    \caption{Reynolds-number folds of the continued Oldroyd--B
    neutral-saddle boundaries.}
    \label{tab:film_fold_summary}
    \begin{tabular}{cccccc}
        \toprule
        $\theta$
        & $Re_F$
        & $De_F$
        & $k_{r,F}$
        & $k_{i,F}$
        & $\omega_{r,F}$
        \\
        \midrule
        $\frac{\pi}{32}$
        & 36.515978
        & 0.499012
        & 1.094245
        & $-0.743253$
        & 0.818055
        \\
        $\frac{\pi}{16}$
        & 17.704969
        & 0.759318
        & 0.710518
        & $-0.426186$
        & 0.657451
        \\
        $\frac{\pi}{12}$
        & 11.435099
        & 0.897633
        & 0.612160
        & $-0.372661$
        & 0.632638
        \\
        $\frac{\pi}{8}$
        & 4.949590
        & 1.137316
        & 0.510762
        & $-0.310345$
        & 0.635903
        \\
        \bottomrule
    \end{tabular}
\end{table}

The four production calculations contain 528 accepted boundary points in
total. 
The mean number of Newton iterations is 2.25 per point,
and no corrector required more than three iterations. Over all four branches,
the maximum direct, full-adjoint, saddle, neutrality, and biorthogonality
residuals are
\begin{align}
    \max\|\mathbf{L}_f\mathbf{q}_f\|_2
    &=
    9.24\times10^{-11}, \notag
    \\
    \max\|\mathbf{L}_f^H\mathbf{q}_f^\dagger\|_2
    &=
    9.80\times10^{-9},\notag
    \\
    \max
    \left|
    (\mathbf{q}_f^\dagger)^H
    \mathbf{C}_f\mathbf{q}_f
    \right|
    &=
    1.38\times10^{-10},\notag
    \\
    \max|\omega_i|
    &=
    7.65\times10^{-27},\notag
    \\
    \max
    \left|
    (\mathbf{q}_f^\dagger)^H
    \mathbf{B}_f\mathbf{q}_f-1
    \right|
    &=
    1.73\times10^{-12}.\notag
\end{align}

The Gaussian-wake calculation does not cross a true fold (turning point), while
the film problem provides the corresponding genuine turning-point test. We use the
$\theta=\frac{\pi}{16}$ boundary as the detailed case because it is the baseline
inclination used in the resolution study.
Since $De$ remains monotone through the fold, the turning point is refined
by repeated fixed-$De$ neutral-saddle corrections and a safeguarded root
search for $t_{Re}^{(p)}(De)=0.$
At the best fold-refinement iterate,
$|t_{Re}^{(p)}| = 6.73\times10^{-9},$
$ t_{De}^{(p)}\simeq1.$
A local fit to the natural-continuation points nearest the fold gives
$Re_F-Re \propto |De-De_F|^{2.008},$
consistent with the quadratic geometry of a simple fold.
As shown in Fig.  \ref{fig:film_boundary_four_theta}(b), 
the natural continuation in $Re$ is also tested for $\theta=\frac{\pi}{16}$.
Starting from the pre-fold branch, fixed-$Re$ neutral-saddle corrections are
performed with a secant predictor and adaptive $\Delta Re$. 
The adaptive Reynolds-number increment decreases from
$O(10^{-2})$ to the imposed lower limit $O(10^{-5})$, and the calculation
terminates after repeated failures of targets above $Re_F$. 
Pseudo-arclength continuation crosses the Reynolds-number fold,
whereas natural continuation in $Re$ develops a collapsing fixed-parameter singular value and repeated corrector failures.
The film example therefore provides a numerical demonstration of the regularising role of the
pseudo-arclength condition when a physical continuation parameter loses local
uniqueness.

\subsection{Effect of elasticity and inclination on the AI/CI transition}
\label{subsec:film_physics_results}

The identification of the smaller-$Re$ side as absolutely unstable gives the
boundary geometry a direct physical interpretation. Starting from $De=0$,
the critical Reynolds number initially increases with $De$ and reaches
$Re_F$ at the fold, see Fig.~\ref{fig:film_boundary_four_theta}(a). 
The absolutely unstable region therefore expands toward
larger Reynolds numbers as weak-to-moderate elasticity is introduced. Beyond
the fold, $Re_c$ decreases with further increase of $De$, so the trend
reverses and the absolutely unstable region contracts.

The fold also implies a re-entrant transition at fixed Reynolds number.
For a given inclination and $Re_c(0)<Re<Re_F,$
a vertical line in the $(Re,De)$ plane intersects the connected
neutral-saddle branch twice. Because the smaller-$Re$ side is absolutely
unstable, increasing $De$ at fixed $Re$ produces CI--AI--CI transition.
Elasticity therefore does not act as a uniformly stabilising or
destabilising parameter in the spatio-temporal problem. Instead, it first
promotes the absolute mode over the pre-fold segment and suppresses it again
on the post-fold segment.
The present calculations establish the resulting shift
of the selected neutral saddle and the re-entrant AI/CI geometry. A detailed
energetic or stress-budget decomposition of the underlying viscoelastic
mechanism is outside the scope of the continuation study and would require a
separate modal-energy analysis.

\section{Conclusions}
\label{sec:conclusions}

A direct adjoint-augmented continuation framework has been developed for
tracking neutral stationary-saddle boundaries in frequency-affine
generalised eigenvalue problems. The central idea is to embed the
zero-group-velocity condition in the nonlinear eigenvalue system through an
adjoint solvability residual. The direct and adjoint eigenproblems, the
stationarity condition, the eigenvector gauges and the neutral-growth
condition are thereby solved as a single augmented system. When two physical
parameters are allowed to vary, the neutral saddles are represented directly
as a one-dimensional solution manifold and followed by scaled
pseudo-arclength continuation.

This formulation changes the numerical object being computed. Conventional
boundary reconstruction commonly requires a saddle search over the complex
wavenumber plane at many parameter values, followed by an additional search
or contour extraction in the physical parameter plane. The present approach
instead advances directly along the neutral-saddle manifold. The
Gaussian-wake calculations show that the adjoint formulation is consistent
with a separately formulated finite-difference saddle residual and recovers
the spatial pinch topology of the selected mode. The comparison with nested
saddle scanning further demonstrates the computational consequence of the
manifold formulation: increasing the resolution of a parameter-plane scan
improves the reconstructed boundary only by repeating the underlying
spectral and saddle calculations at additional parameter points, whereas
pseudo-arclength continuation concentrates the computation on the boundary
itself. Across the tested scan resolutions, the nested procedure requires
$8.1$--$52.2$ times the direct-continuation wall time; moreover,
extrapolating the measured $E_H$--$T$ scaling to
$E_H\sim10^{-6}$ suggests an estimated cost ratio of approximately
$1.8\times10^{4}$ in favour of the direct adjoint continuation.

The Oldroyd--B film problem demonstrates that the same framework extends to
a coupled multi-field free-surface eigenproblem without a problem-specific
continuation reformulation. The continued neutral-saddle manifolds develop
genuine folds in their Reynolds-number projections. Near such a fold,
natural continuation in the Reynolds number loses local solvability, whereas
pseudo-arclength continuation remains regular and resolves the post-fold
branch. For the selected saddle family examined here, the resulting
$(Re,De)$ boundary geometry shows that elasticity does not shift the
neutral boundary monotonically: the pre- and post-fold branches produce a
re-entrant CI--AI--CI sequence over part of the parameter range. The film
application therefore illustrates both the numerical role of the
pseudo-arclength constraint and the additional boundary structure that may
be missed when a single physical parameter is imposed as the continuation
coordinate.

The present method is a correction and continuation framework for a selected
saddle family. It assumes a locally simple finite temporal eigenvalue and a
regular one-dimensional augmented solution manifold, and it requires an
approximate initial saddle and modal-family identity. Moreover, the
algebraic conditions
$\mathrm{d}\omega/\mathrm{d} k=0$ and $\operatorname{Im}\omega=0$ identify neutral
stationary eigenpairs but do not by themselves establish Briggs--Bers
pinching or dominance over competing saddle families. Spatial-branch
topology and possible changes in saddle dominance must therefore be examined
separately when required by the physical interpretation.

Subject to these qualifications, the results establish adjoint-augmented
pseudo-arclength continuation as a direct route for computing connected
neutral-saddle manifolds in multi-parameter spatio-temporal stability
problems. Its principal advantage is not merely a different evaluation of
the eigenvalue derivative, but the replacement of nested saddle searches and
parameter-plane boundary reconstruction by direct continuation of the
boundary as a solution manifold. This viewpoint provides a natural
framework for extending neutral-saddle calculations to larger coupled
eigenvalue systems, parameter folds and more complex multi-parameter
stability boundaries.

\appendix

\section{Gaussian-wake spatial-discretisation convergence}
\label{app:wake_convergence}

This appendix documents the spatial-discretisation checks used to select the
production grid. These tests are kept
separate from the adjoint-method verification in Sec.~\ref{sec:wake} because
they assess the mapped Orr--Sommerfeld discretisation rather than the
solvability identity, the semi-analytical Jacobian, or the continuation
algorithm.

\subsection{Discretisation homotopy and collocation-order convergence}
\label{app:wake_N_convergence}

The neutral saddle at $\lambda=0.95$ is first corrected on a coarse mapped
grid with $N=60$ and then transferred through the discretisation homotopy
\begin{equation}
    N=60\rightarrow70\rightarrow80\rightarrow90\rightarrow100.
    \label{eq:app_wake_N_homotopy}
\end{equation}
At each step, the direct mode on the old physical grid is reconstructed,
interpolated to the new mapped grid, and used together with eigenvalue
proximity and modal overlap to identify the corresponding temporal mode. The
full direct--adjoint neutral-saddle system is then corrected at the new
resolution. The same mode-transfer procedure is used in the convergence
sweeps to avoid interpreting convergence to different discrete saddle
families as a resolution trend.

Table~\ref{tab:wake_N_convergence} reports the collocation-order study at
fixed $y_{\max}=20$ and $\beta=3$. The coarse calculations are visibly
under-resolved, whereas $N\ge90$ approaches a clear plateau. Relative to the
$N=120$ result, the production-grid value at $N=100$ differs by
$1.263\times10^{-7}$ in relative $Re_c$. The absolute differences in
$\alpha_s$ and $\omega_s$ are $8.679\times10^{-8}$ and
$1.672\times10^{-9}$, respectively.

\begin{table}[htbp]
    \centering
    \caption{Collocation-order convergence of the sinh-mapped Gaussian-wake
    neutral saddle at $\lambda=0.95$, $y_{\max}=20$, and $\beta=3$.}
    \label{tab:wake_N_convergence}
    \begin{tabular}{rrrrrr}
        \toprule
        $N$ &
        $\Delta y_0$ &
        $Re_c$ &
        $\alpha_r$ &
        $\alpha_i$ &
        $\omega_r$
        \\
        \midrule
         50 & 0.3783 & 323.240030 & 0.938397 & -0.859683 & 0.525732 \\
         60 & 0.3147 & 336.162021 & 0.933571 & -0.862114 & 0.526014 \\
         70 & 0.2695 & 333.067718 & 0.934253 & -0.862605 & 0.526000 \\
         80 & 0.2357 & 333.214491 & 0.934281 & -0.862462 & 0.525996 \\
         90 & 0.2094 & 333.237462 & 0.934266 & -0.862469 & 0.525997 \\
        100 & 0.1884 & 333.234821 & 0.934267 & -0.862470 & 0.525997 \\
        110 & 0.1712 & 333.234858 & 0.934267 & -0.862470 & 0.525997 \\
        120 & 0.1569 & 333.234863 & 0.934267 & -0.862470 & 0.525997 \\
        140 & 0.1345 & 333.234863 & 0.934267 & -0.862470 & 0.525997 \\
        \bottomrule
    \end{tabular}
\end{table}

\subsection{Mapping-parameter convergence}
\label{app:wake_beta_convergence}

The sinh-mapping parameter is varied from $\beta=2$ to $4$ at fixed
$N=100$ and $y_{\max}=20$. The values at $\beta=3$ and $3.5$ differ by
only $1.257\times10^{-7}$ in relative $Re_c$ and by
$8.672\times10^{-8}$ in the complex saddle wavenumber. The results at
$\beta=3.5$ and $4$ are indistinguishable to the reported digits. Thus, once
the wake core is sufficiently resolved, the neutral saddle is not controlled
by the arbitrary mapping parameter.

\subsection{Domain-truncation convergence}
\label{app:wake_ymax_convergence}

A separate domain-truncation study varied $y_{\max}$ while choosing an even
$N$ for each domain so that the central physical spacing remained
approximately equal to that of the production grid. The pairs
\begin{equation}
    (y_{\max},N)
    =
    (10,50),\,
    (15,76),\,
    (20,100),\,
    (25,124),\,
    (30,150)
    \label{eq:app_wake_ymax_pairs}
\end{equation}
give $0.186\lesssim\Delta y_0\lesssim0.190$, with a maximum relative
wake-core spacing mismatch of approximately $1.22\%$. The resulting critical
Reynolds numbers are $333.229524$, $333.234632$, $333.234821$,
$333.234916$, and $333.234912$, respectively. The relative difference in
$Re_c$ between $y_{\max}=20$ and $30$ is $2.742\times10^{-7}$; the
corresponding absolute differences in $\alpha_s$ and $\omega_s$ are
$5.198\times10^{-8}$ and $1.475\times10^{-10}$.

\begin{figure}[htbp]
    \centering
    \includegraphics[width=1\textwidth]{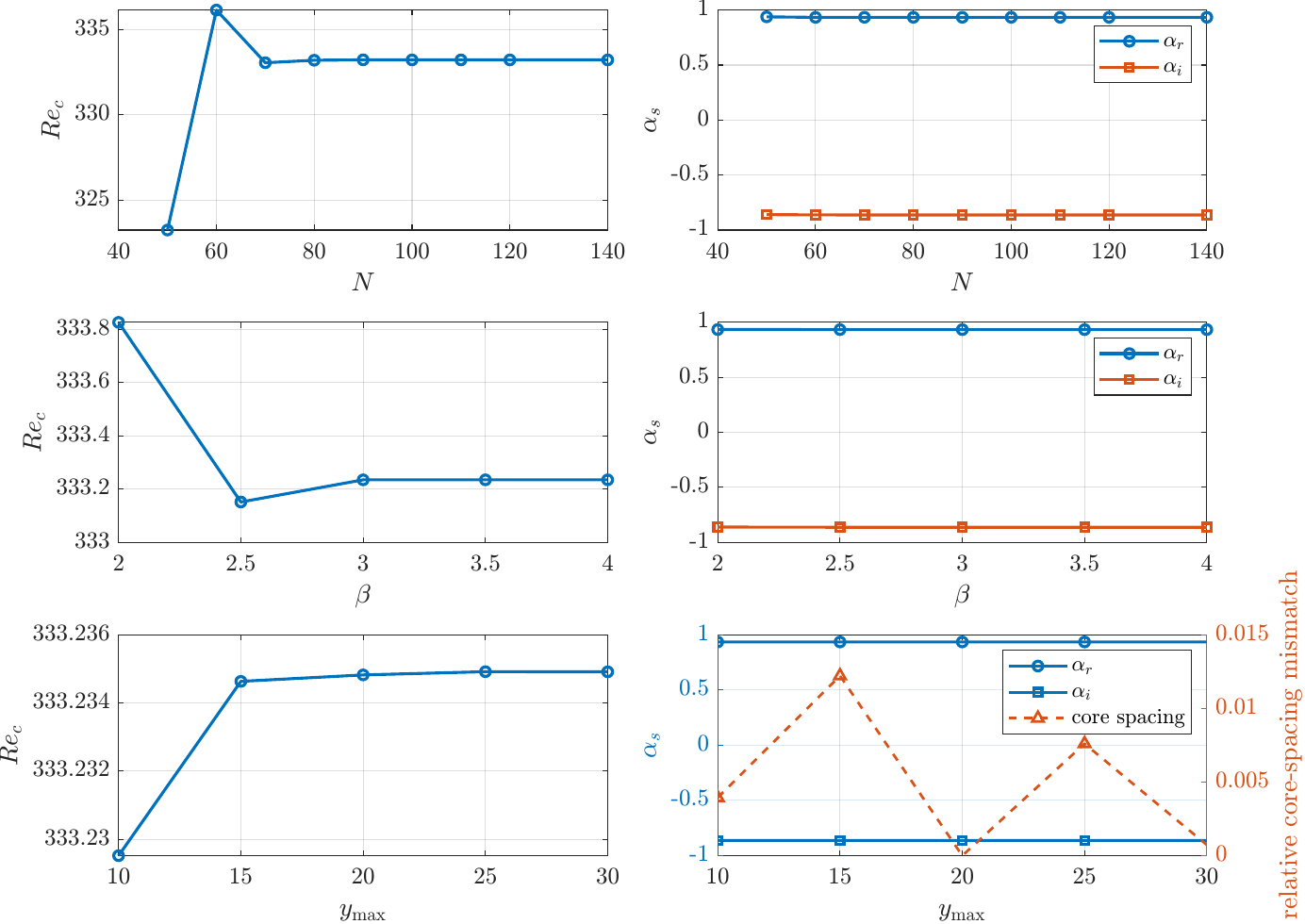}
    \caption{Convergence of the Gaussian-wake neutral saddle with respect to
    the collocation order $N$, sinh-mapping parameter $\beta$, and
    truncated-domain half-width $y_{\max}$. In the $y_{\max}$ sweep, $N$ is
    adjusted to keep the wake-core spacing approximately fixed.}
    \label{fig:wake_convergence}
\end{figure}

Figure~\ref{fig:wake_convergence} summarises the three studies. The
collocation-order, mapping-parameter, and domain-truncation results jointly
support the parameters of $N=100$, $y_{\max}=20$, and $\beta=3$.

\section*{Author Contributions}

Z. Ding conceived the algorithm and supervised the research.
Y. Xiao implemented the algorithm, performed the data analysis, and prepared the original manuscript. H. Li and Z. Ding contributed to the theoretical
discussion, interpretation of the results, and revision of the manuscript.
All authors reviewed and approved the final version of the manuscript.

\bibliographystyle{siamplain}
\bibliography{references}
\end{document}